\documentclass[12pt]{article}

\usepackage{cite}                

\usepackage{amssymb}             

\usepackage{latexsym}            

\usepackage{amsfonts}            

\usepackage{amsmath}             

\usepackage{stmaryrd}            

\usepackage[mathscr]{eucal}

\input xypic                     

\newtheorem{The}{Theorem}

\newtheorem{Lem}{Lema}[section]
\newtheorem{Prop}{Proposition}[section]
\newtheorem{Cor}{Corollary}[section]

\newcommand{\bthe}{\begin{The}}
\newcommand{\ethe}{\end{The}}
\newcommand{\blem}{\begin{Lem}}
\newcommand{\elem}{\end{Lem}}
\newcommand{\bpro}{\textsc{Proof:} \rm}
\newcommand{\epro}{\hfill $\Box$}
\newcommand{\bprop}{\begin{Prop}}
\newcommand{\eprop}{\end{Prop}}
\newcommand{\bcor}{\begin{Cor}}
\newcommand{\ecor}{\end{Cor}}

\newcommand{\bdefi}{\begin{defi} \rm}
\newcommand{\edefi}{\end{defi}}
\newcommand{\bobs}{\begin{obs} \rm}
\newcommand{\eobs}{\end{obs}}

\newcommand{\R}{\ensuremath{\mathbb{R}}}

\newcommand{\Map}{{\rm Map}}

\newcommand{\mathslf}[1]{\ensuremath{\mbox{\slshape \textsf{#1}}}}
\newcommand{\mathsmf}[1]{\ensuremath{\mbox{\scriptsize \slshape \textsf{#1}}}}

\newcommand{\smin}{\,\raisebox{0.06em}{${\scriptstyle \in}$}\,}
\newcommand{\nsmin}{\,\raisebox{0.06em}{${\scriptstyle \notin}$}\,}

\newcommand{\smsubset}{\,\raisebox{0.06em}{${\scriptstyle \subset}$}\,}

\newcommand{\smcup}{\,\raisebox{0.06em}{${\scriptstyle \cup}$}\,}

\newcommand{\smwedge}{{\scriptstyle \wedge}}

\newcommand{\smcirc}{\,\raisebox{0.1ex}{${\scriptstyle \circ}$}\,}
\newcommand{\bvee}[1]{\raisebox{0.2ex}{${\textstyle \bigvee}$}
 \ensuremath{^{\raisebox{-0.2ex}{${\scriptstyle #1}$}}\,}}
\newcommand{\bwedge}[1]{\raisebox{0.2ex}{${\textstyle \bigwedge}$}
 \ensuremath{^{\raisebox{-0.2ex}{${\scriptstyle #1}$}}\,}}

\newcommand{\oast}{\ensuremath{\:\!%
 {\raisebox{0.1ex}{$\scriptscriptstyle \bigcirc$} \hspace{-0.5em} \ast}}}
\newcommand{\ostar}{\ensuremath{\:\!%
 {\raisebox{0.05ex}{$\scriptscriptstyle \bigcirc$} \hspace{-0.5em} \star}}}

\newcommand{\vect}[1]{\ensuremath{\vec{\:\!#1}\:\!}}

\begin{document}

\title{Covariant Poisson Brackets \\
       in Geometric Field Theory}
\author{
  Michael Forger$\,^1\,$\thanks{Partially supported by CNPq, Brazil}~
  and
  Sandro Vieira Romero$\,^2\,$\thanks{Partially supported by FAPESP, Brazil}}
\date{\normalsize
      $^1\,$ Departamento de Matem\'atica Aplicada,~~\mbox{} \\
      Instituto de Matem\'atica e Estat\'{\i}stica, \\
      Universidade de S\~ao Paulo, \\
      Caixa Postal 66281, \\
      BR--05311-970~ S\~ao Paulo SP, Brazil \\[4mm]
      $^2\,$ Departamento de Matem\'atica,~~\mbox{} \\
      Universidade Federal de Vi\c{c}osa,\\
      BR--36571-000~ Vi\c{c}osa MG, Brazil}

\footnotetext[1]{\emph{E-mail address:} \textsf{forger@ime.usp.br}}
\footnotetext[2]{\emph{E-mail address:} \textsf{sromero@ufv.br}}
\maketitle

\thispagestyle{empty}

\begin{abstract}
\noindent
We establish a link between the multisymplectic and the covariant phase space
approach to geometric field theory by showing how to derive the symplectic
form on the latter, as introduced by Crnkovi\'c-Witten and Zuckerman, from
the multi\-symplectic form. The main result is that the Poisson bracket
associated with this symplectic structure, according to the standard
rules, is precisely the covariant bracket due to Peierls and DeWitt.
\end{abstract}
\vspace{-5mm}
\begin{flushright}
 \parbox{12em}
 {\begin{center}
   Universidade de S\~ao Paulo \\
   RT-MAP\,-\,0304 \\
   December 2003
  \end{center}}
\end{flushright}

\newpage

\section{Introduction}

One of the most annoying flaws of the usual canonical formalism in field
theory is its lack of manifest covariance, that is, its lack of explicit
Lorentz invariance (in the context of special relativity) and more generally
its lack of explicit invariance under space-time coordinate transformations
(in the context of general relativity). Of course, this defect is built
into the theory from the very beginning, since the usual canonical
formalism represents the dynamical variables of classical field theory
by functions on some spacelike hypersurface (Cauchy data) and provides
differential equations for their time evolution off this hypersurface:
thus it presupposes a splitting of space-time into space and time, in
the form of a foliation of space-time into Cauchy surfaces. As a result,
\linebreak
canonical quantization leads to models of quantum field theory whose
covariance is far from obvious and in fact constitutes a formidable
problem: as a well known example, we may quote the efforts necessary
to check Lorentz invariance in (perturbative) quantum electrodynamics
in the Coulomb gauge.

These and similar observations have over many decades nourished attempts
to develop a fully covariant formulation of the canonical formalism in
classical field theory, which would hopefully serve as a starting point
for alternative methods of quantization. \linebreak Among the many ideas
that have been proposed in this direction, two have come to occupy a
special role. One of these is the ``covariant functional formalism'',
based on the concept of ``covariant phase space'' which is defined as
the (infinite-dimensional) space of solutions of the equations of motion.
This approach was strongly advocated in the 1980's by Crnkovi\'c, Witten
and Zuckerman~\cite{CW,Cr,Zu} (see also~\cite{Wo}) who showed how to
construct a symplectic structure on the covariant phase space of many
important models of field theory (including gauge theories and general
relativity), but the idea as~such has a much longer history. The other
has become known as the ``multisymplectic \linebreak formalism'', based
on the concept of ``multiphase space'' which is a (finite-dimensional)
space that can be defined locally by associating to each coordinate $q^i$
not just one conjugate momentum $p\>\!_i$ but $n$ conjugate momenta
$p\>\!_i^\mu$ ($\mu = 1,\ldots,n$), where $n$ is the dimension of the
underlying space-time manifold. In coordinate form, this construction
goes back to the classical work of De Donder and Weyl in the 1930's
\cite{DD,We}, whereas a global formulation was initiated in the 1970's
by a group of mathematical physicists, mainly in Poland~\cite{Ki,KS1,KS2}
but also elsewhere~\cite{GoS,GuS,Gar}, and definitely established in the
1990's~\cite{CCI,Go}; a detailed exposition, with lots of examples, can
be found in the GIMmsy paper~\cite{GIM}.

The two formalisms, although both fully covariant and directed towards
the same ulti\-mate goal, are of different nature; each of them has its
own merits and drawbacks.
\begin{itemize}
 \item The multisymplectic formalism is manifestly consistent with the
       basic principles of field theory, preserving full covariance, and
       it is mathematically rigorous because it uses well established
       methods from calculus on finite-dimensional manifolds. On~the
       other hand, it does not seem to permit any obvious definition
       of the \mbox{Poisson} bracket between observables. Even the
       question of what mathematical objects should represent physical
       observables is not totally clear and has in fact been the subject
       of much debate in the literature. Moreover, the introduction of
       $n$ conjugate momenta for each coordinate obscures the usual
       duality between canonically conjugate variables (such as momenta
       and positions), which plays a fundamental role in all known methods
       of quantization. A definite solution to these problems has yet to be
       found.
 \item The covariant functional formalism fits neatly into the philosophy
       underlying the symplectic formalism in general; in particular, it
       admits a natural definition of the Poisson bracket (due to Peierls
       \cite{Pe} and further elaborated by DeWitt \cite{DW1,DW2,DW3})
       \linebreak that preserves the duality between canonically conjugate
       variables. Its main drawback is the lack of mathematical rigor, since
       it is often restricted to the formal extrapolation of techniques from
       ordinary calculus on manifolds to the infinite-dimensional setting:
       transforming such formal results into mathematical theorems is a
       separate problem, often highly complex and difficult.
\end{itemize}
Of course, the two approaches are closely related, and this relation
has been an \linebreak important source of motivation in the early days
of the theory \cite{KS1}. Unfortunately, however, the tradition of
developing them in parallel seems to have partly fallen into oblivion
in recent years, during which important progress was made in other
directions. 

The present paper, based on the PhD thesis of the second author \cite{Rom}, is
intended to revitalize this tradition by systematizing and further developing
the link between the two approaches, thus contributing to integrate them into
one common picture. It~is~organized into two main sections. In Sect.~2, we
briefly review some salient features of the multisymplectic approach to
geometric field theory, focussing on the concepts needed to make contact
with the covariant functional approach. In particular, this requires a
digression on jet bundles of first and second order as well as on the
definition of both extended and ordinary multiphase space as the twisted
affine dual of the first order jet bundle and the twisted linear dual of
the linear first order jet bundle, respectively: this will enable us to
give a global definition of the space of solutions of the equations of
motion, both in the Lagrangian and Hamiltonian formulation, in terms
of a globally defined Euler\,-\,Lagrange operator~$\mathscr{E}$ and a
globally defined De Donder\,-\,Weyl operator~$\mathscr{D}$, respectively.
To describe the formal tangent space to this space of solutions at a given
point, we also write down the linearization of each of these operators
around a given solution. In~Sect.~3, we apply these constructions to
derive a general expression for the symplectic form $\Omega$ on covariant
phase space, \`a la Crnkovi\'c-Witten-Zuckerman, in terms of the multi%
symplectic form $\,\omega$ on extended multiphase space. Then we prove,
as the main result of this paper, that the Poisson bracket associated
with the form~$\Omega$, according to the standard rules of symplectic
geometry, suitably extended to this infinite-dimensional setting, is
precisely the Peierls\,-\,DeWitt bracket of classical field theory
\cite{Pe,DW1,DW2,DW3}. Finally, in Sect.~4, we comment on the relation
of our results to previous work and on perspectives for future research
in this area.

\section{Multisymplectic Approach}

\subsection{Overview}

The multisymplectic approach to geometric field theory, whose origins
can be traced back to the early work of Hermann Weyl on the calculus of
variations \cite{We}, is based on the idea of modifying the transition
from the Lagrangian to the Hamiltonian framework by treating spatial
derivatives and time derivatives of fields on an equal footing. Thus
one associates to each field component $\varphi^{\,i}$ not just its
standard canonically conjugate momentum $\pi_i^{}$ but rather $n$
conjugate momenta $\pi\:\!_i^\mu$, where $n$ is the dimension of
space-time. In a first order Lagrangian formalism, where one starts
out from a~Lagrangian $L$ depending on the field and its first partial
derivatives, these are obtained by a covariant analogue of the Legendre
transformation
\begin{equation} \label{eq:CCMM1}
 \pi_i^\mu~
 =~\frac{\partial L}{\partial \, \partial_\mu \varphi^{\,i}}~.
\end{equation}
This allows to rewrite the standard Euler\,-\,Lagrange equations of
field theory
\begin{equation} \label{eq:ELEQ1}
 \partial_\mu \,
 \frac{\partial L}{\partial \, \partial_\mu \varphi^i} \, - \,
 \frac{\partial L}{\partial \varphi^i}~=~0
\end{equation}
as a covariant first order system, the covariant Hamiltonian equations or
De Donder\,-\,Weyl equations
\begin{equation} \label{eq:DWEQ1}
 \frac{\partial H}{\partial \pi_i^\mu}~
 =~\partial_\mu \varphi^i \quad , \quad
 \frac{\partial H}{\partial \varphi^i}~
 = \;- \, \partial_\mu \pi_i^\mu
\end{equation}
where
\begin{equation} \label{eq:DWHAM}
 H~=~\pi_i^\mu \, \partial_\mu \varphi^i \, - \, L
\end{equation}
is the covariant Hamiltonian density or De Donder\,-\,Weyl Hamiltonian.

Multiphase space (ordinary as well as extended) is the geometric environment
built by appropriately patching together local coordinate systems of the form
$(q^i,p\>\!_i^\mu)$~-- in- \linebreak stead of the canonically conjugate
variables $(q^i,p_i)$ of mechanics~-- together with space-time coordinates
$x^\mu$ and, in the extended version, a further energy type variable that
we shall denote by $p$ (without any index). The global construction of
these multiphase spaces, however, has only gradually come to light; it
is based on the following mathematical concepts.
\begin{itemize}
 \item The collection of all fields in a given theory, defined over a fixed
       ($n$-dimensional orientable) space-time manifold $M$, is represented
       by the sections $\varphi$ of a given fiber bundle $E$ over $M$, with
       bundle projection $\, \pi : E \rightarrow M \,$ and typical fiber $Q$.
       This bundle will be referred to as the \emph{configuration bundle}
       of the theory since $Q$ corresponds to the configuration space of
       possible field values.
 \item The collection of all fields together with their partial derivatives
       up to a certain order, say order $r$, is represented by the $r$-jets
       $\, j^{\,r\!} \varphi \equiv (\varphi,\partial\varphi,\ldots,
       \partial^{\,r\!} \varphi)$  of sections of $E$, which are
       themselves sections of the $r^{\mathrm{th}}$ order jet bundle
       $J^r E$ of $E$, regarded as a fiber bundle over~$M$. In this paper,
       we shall only need first order jet bundles, \linebreak with one
       notable exception: the global formulation of the Euler\,-\,Lagrange
       equations requires introducing the second order jet bundle.
 \item Dualization -- the concept needed to pass from the Lagrangian to
       the Hamiltonian framework via the Legendre transformation -- comes
       in two variants, based on the fundamental observation that the first
       order jet bundle $J^1 E$ of $E$ is an affine bundle over $E$ whose
       difference vector bundle $\vec{J}^{\,1} E$ will be referred to as
       the linear jet bundle. Ordinary multiphase space is obtained as
       the twisted linear dual $\vec{J}^{\,1\oast} E$ of $\vec{J}^{\,1} E$
       while extended multiphase space is obtained as the twisted affine
       dual $J^{1\ostar} E$ of $J^1 E$, where the prefix ``twisted'' refers
       to the necessity of taking an additional tensor product with the
       bundle of $n$-forms on $M$.\footnote{We use an asterisk $\ast$ to
       denote linear duals of vector spaces or bundles and a star $\star$
       to denote affine duals of affine spaces or bundles. These symbols
       are appropriately encircled to characterize twisted duals, as
       opposed to the ordinary duals defined in terms of linear or
       affine maps with values in $\mathbb{R}$.}
 \item The Lagrangian $\mathscr{L}$ is a function on $J^1 E$ with values
       in the bundle of $n$-forms on~$M$ so that it may be integrated to
       provide an action functional which enters the variational principle.
       The De Donder\,-\,Weyl Hamiltonian $\mathscr{H}$ is a section of
       $J^{1\ostar} E$, considered as an affine line bundle over
       $\vec{J}^{\,1\oast} E$.
\end{itemize}
Note that the formalism is set up so as to require no additional structure
on the con\-figuration bundle or on any other bundle constructed from it:
all are merely fiber bundles over the space-time manifold $M$. Of course,
additional structures do arise when one is dealing with special classes
of fields (matter fields and the metric tensor in general relativity are
sections of vector bundles, connections are sections of affine bundles,
nonlinear fields such as those arising in the sigma model are sections
of trivial fiber bundles with a fixed Riemannian metric on the fibers,
etc.), but such additional structures depend on the kind of theory
considered and thus are not universal. Finally, the restriction
imposed on the order of the jet bundles considered reflects the
fact that almost all known examples of field theories are governed by
second order partial differential equations which can be derived from a
Lagrangian that depends only on the fields and their partial derivatives
of first order, which is why it is reasonable to develop the general
theory on the basis of a first order formalism, as is done in mechanics
\cite{AM,Arn}.

\subsection{The First Order Jet Bundle}

The field theoretical analogue of the tangent bundle of mechanics is the
\emph{first order jet bundle} $J^1 E$ associated with the configuration
bundle $E$ over $M$. Given a point $e$ in~$E$ with base point $\, x =
\pi(e) \,$ in~$M$, the fiber $J_e^1 E$ of~$J^1 E$ at~$e$ consists of all
linear maps from the tangent space $T_x^{} M$ of the base space~$M$ at~$x$
to the tangent space $T_e^{} E$ of the total space~$E$ at~$e$ whose compo%
sition with the tangent map $\, T_e^{} \pi: T_e^{} E \rightarrow T_x^{} M \,$
to the projection $\, \pi: E \rightarrow M \,$ gives the identity
on~$T_x^{} M$:
\begin{equation} \label{eq:FJB1}
 J_e^1 E~=~\{ \, \gamma \in L(T_x^{} M,T_e^{} E) \,:\,
              T_e^{} \pi \circ \gamma~=~\mathrm{id}_{T_x^{} M}^{} \, \}~.
\end{equation}
Thus the elements of $J_e^1 E$ are precisely the candidates for the tangent
maps at~$x$ to (local) sections $\varphi$ of the bundle~$E$ satisfying 
$\, \varphi(x) = e$. Obviously, $J_e^1 E$ is an affine subspace of the
vector space $\, L(T_x^{} M, T_e^{} E) \,$ of all linear maps from~$T_x^{} M$
to the tangent space $T_e^{} E$, the corresponding difference vector space
being the vector space of all linear maps from~$T_x^{} M$ to the vertical
subspace~$V_e^{} E$:
\begin{equation} \label{eq:LJB1}
 \vec{J}_{\,e}^{\,1} E~
 =~\{ \, \vect{\gamma} \in L(T_x^{} M,T_e^{} E) \,:\,
         T_e^{} \pi \circ \vect{\gamma}~=~0 \, \}~
 =~L(T_x^{} M,V_e^{} E)~
 \cong~T_x^\ast M \otimes V_e^{} E~.
\end{equation}
The jet bundle $J^1E$ thus defined admits two different projections, namely
the \emph{target projection} $\, \tau_E^{} : J^1 E \rightarrow E \,$ and the
\emph{source projection} $\, \sigma_E^{} : J^1 E \rightarrow M \,$ which is
simply its composition with the original bundle projection, that~is,
$\sigma_E^{} = \pi \circ \tau_E^{}$. The same goes for $\vec{J}^{\,1} E$,
which we shall call the linearized first order jet bundle or simply
\emph{linear jet bundle} associated with the configuration bundle~$E$
over~$M$.

The structure of $J^1 E$ and of $\vec{J}^{\,1} E$ as fiber bundles over~$M$
with respect to the source projection (in general without any additional
structure), as well as that of $J^1 E$ as an affine bundle and of~%
$\vec{J}^{\,1} E$ as a vector bundle over~$E$ with respect to the target
projection, can most easily be seen in terms of local coordinates. Namely,
local coordinates $x^\mu$ for~$M$ and $q^i$ for~$Q$, together with a local
trivialization of~$E$, induce local coordinates $(x^\mu,q^i)$ for~$E$ as
well as local coordinates $(x^\mu,q^i,q_\mu^i)$ for $J^1 E \smsubset
L(\pi^\ast(TM),TE)$ and $(x^\mu,q^i,\vect{q}_\mu^{\,i})$ for
$\vec{J}^{\,1} E \smsubset L(\pi^\ast(TM),TE)$. Moreover, local
coordinate transformations $\, x^\mu \rightarrow x^{\prime\nu} \,$
for~$M$ and $\, q^i \rightarrow q^{\prime j} \,$ for~$Q$, together
with a change of local trivialization of~$E$, correspond to a local
coordinate transformation $\, (x^\mu,q^i) \rightarrow (x^{\prime\nu},
q^{\prime j}) \,$ for~$E$ where
\begin{equation} \label{eq:LCTR1}
 x^{\prime\nu}~=~x^{\prime\nu}(x^\mu)~~~,~~~
 q^{\prime j}~=~q^{\prime j}( x^\mu,q^i )~.
\end{equation}
The induced local coordinate transformations $\, (x^\mu,q^i,q_\mu^i)
\rightarrow (x^{\prime\nu},q^{\prime j},q_\nu^{\prime j}) \,$ for $J^1 E$
and $\, (x^\mu,q^i,\vect{q}_\mu^{\,i}) \rightarrow (x^{\prime\nu},
q^{\prime j},\vect{q}_\nu^{\,\prime j}) \,$ for $\vec{J}^{\,1} E$
are then easily seen to be given by
\begin{equation} \label{eq:LCTR2}
 q_\nu^{\prime j}~
 =~\frac{\partial x^\mu}{\partial x^{\prime\nu}} \;
   \frac{\partial q^{\prime j}}{\partial q^i} \; q_\mu^i \; + \;
   \frac{\partial x^\mu}{\partial x^{\prime\nu}} \;
   \frac{\partial q^{\prime j}}{\partial x^\mu}~,
\end{equation}
and
\begin{equation} \label{eq:LCTR3}
 \vect{q}_\nu^{\,\prime j}~
 =~\frac{\partial x^\mu}{\partial x^{\prime\nu}} \;
   \frac{\partial q^{\prime j}}{\partial q^i} \; \vect{q}_\mu^{\,i}~.
\end{equation}

\pagebreak

\noindent
This makes it clear that $J^1 E$ is an affine bundle over $E$ with difference
vector bundle
\begin{equation}
 \vec{J}^{\,1} E~=~T^\ast M \otimes V\!E~,
\end{equation}
in accordance with eq.~(\ref{eq:LJB1}).\footnote{Given any vector bundle
$V$ over $M$, such as $TM$, $T^\ast M$ or any of their exterior powers,
one can consider it as as vector bundle over $E$ by forming its pull-back
$\pi^\ast V$. In order not to overload the notation, we shall here and in
what follows suppress the symbol $\pi^\ast$.}

That the (first order) jet bundle of a fiber bundle is the adequate arena to
incorporate (first order) derivatives of fields becomes apparent by noting that
a global section $\varphi$ of~$E$ over~$M$ naturally induces a global section 
$j^1 \varphi$ of~$J^1 E$ over~$M$ given by
\[
 j^1 \varphi(x)~=~T_x^{} \varphi~\smin~J_{\varphi(x)}^1 E \qquad
 \mbox{for $\, x \smin M$}~.
\]
In the mathematical literature, $j^1 \varphi$ is called the (first)
prolongation of $\varphi$, but it would be more intuitive to simply
call it the derivative of $\varphi$ since in the local coordinates
used above,
\[
 j^1 \varphi(x)~=~(x^\mu,\varphi^i(x),\partial_\mu \varphi^i(x))~,
\]
where $\partial_\mu = \partial / \partial x^\mu$; this is symbolically
summarized by writing $\, j^1 \varphi \equiv (\varphi,\partial\varphi)$.

Similarly, it can be shown that the linear jet bundle of a fiber bundle
is the adequate arena to incorporate covariant derivatives of sections,
with respect to an arbitrarily chosen connection.

Finally, let us discuss briefly the lifting, from $E$ to $J^1 E$, of (local)
bundle automorphisms and, passing to generators of one-parameter groups,
of projectable vector fields. Let $\, \Phi: E \rightarrow E \,$ be an auto%
morphism of the fiber bundle $E$ over $M$ and $\, \phi: M \rightarrow M \,$
the induced diffeomorphism of $M$ such that the diagram
\[
 \begin{array}{rcccl}
                &          E      &
  \stackrel{\rule[-2mm]{0mm}{2mm} {\textstyle \Phi}
            \rule[-2mm]{0mm}{2mm}}{\longrightarrow} & E & \\[3mm]
  \pi \!\!\!\!& \bigg\downarrow & & \bigg\downarrow &\!\!\!\! \pi \\
                &          M      &
  \stackrel{\rule[-2mm]{0mm}{2mm} {\textstyle \phi}
            \rule[-2mm]{0mm}{2mm}}{\longrightarrow} & M &
 \end{array}
\]
commutes. This can be lifted to an automorphism of the jet bundle $J^1 E$,
as an affine bundle over $E$, by defining $\, J^1 \Phi : J^1 E \rightarrow
J^1 E \,$ as follows: given a point $e$ in $E$ with base point $\, x =
\pi(e) \,$ in $M$ and a $1$-jet $\, \gamma \smin J_e^1 E$, define the
$1$-jet $\, J^1 \Phi(\gamma) \smin J_{\Phi(e)}^1 E \,$ by
\begin{equation} \label{eq:FJB12}
 J^1 \Phi(\gamma)~=~T_e^{} \Phi \circ \gamma \circ (T_x^{} \phi)^{-1}~.
\end{equation}
Obviously, this formula defines a linear map from $L(T_x^{} M,T_e^{} E)$
to $L(T_{\phi(x)} M,T_{\Phi(e)} E)$ that restricts to an affine map from
$J_e^1 E$ to $J_{\Phi(e)}^1 E \,$:
\begin{eqnarray*}
 T_{\Phi(e)} \pi \circ J^1 \Phi (\gamma) \!\!
 &=&\!\! T_{\Phi(e)}^{} \pi \circ T_e^{} \Phi \circ
         \gamma \circ (T_x^{} \phi)^{-1}                                \\
 &=&\!\! T_e^{} \left( \pi \circ \Phi \right) \circ
         \gamma \circ (T_x^{} \phi)^{-1}                                \\
 &=&\!\! T_e^{} \left( \phi \circ \pi \right) \circ
         \gamma \circ (T_x^{} \phi)^{-1}                                \\
 &=&\!\! T_x^{} \phi \circ T_e^{} \pi \circ
         \gamma \circ (T_x^{} \phi)^{-1}                                \\
 &=&\!\! T_x^{} \phi \circ \mathrm{id}_{\,T_x^{} M}^{} \circ
         (T_x^{} \phi)^{-1}                                             \\
 &=&\!\! \mathrm{id}_{\,T_{\phi(x)}^{} M}^{}~.
\end{eqnarray*}
In particular, the diagram
$$
 \begin{array}{rcccl}
                      &        J^1 E       &
  \stackrel{\rule[-2mm]{0mm}{2mm} {\textstyle J^1 \Phi}
            \rule[-2mm]{0mm}{2mm}}{\longrightarrow} & J^1 E &           \\[3mm]
  \tau        \!\!\!\!& \bigg\downarrow &
                      & \bigg\downarrow &\!\!\!\! \tau                  \\
                      &         E       &
  \stackrel{\rule[-2mm]{0mm}{2mm} {\textstyle \Phi}
            \rule[-2mm]{0mm}{2mm}}{\longrightarrow} &  E  & 
 \end{array}
$$
commutes, justifying to call $J^1 \Phi$ the \emph{first prolongation}
of $\Phi$. This construction can be generalized to any fiber bundle map 
$\, \Phi : E \rightarrow F \,$ over a (local) diffeomorphism $\, \phi:
M \rightarrow N \,$, giving an affine bundle map $\, J^1 \Phi:
J^1 E \rightarrow J^1 F \,$ over $\, \Phi : E \rightarrow F$.

Passing to the description of the infinitesimal situation, let us consider
a projectable vector field $V$ on $E$, whose flow is a one-parameter group
of (local) automorphisms of $E$ that can be lifted to a one-parameter group
of local automorphisms of $J^1 E$, generated by a projectable vector field
$J^1 V$ on $J^1 E$: this is then defined to be the prolongation of $V$. Thus
\[
 V~=~\frac{\partial \Phi_\lambda}{\partial \lambda} \Big|_{\lambda=0}
 ~~\Longrightarrow~~
 J^1 V~=~\frac{\partial (J^1 \Phi_\lambda)}{\partial \lambda}
         \Big|_{\lambda=0}~.
\]
In local coordinates as before, we can write
\begin{equation} \label{eq:PVFE}
 V~=~V^\mu \, \frac{\partial}{\partial x^\mu} \; + \;
     V^i \, \frac{\partial}{\partial q^i}
\end{equation}
where $\, V^\mu = V^\mu(x^\nu) \,$ and $\, V^i = V^i(x^\nu,q^j)$,
and since the lifting of bundle automorphisms is described by the
transformation law (\ref{eq:LCTR2}), differentiation with respect
to $\lambda$ gives
\begin{equation} \label{eq:LVFJ}
 J^1 V~=~V^\mu \, \frac{\partial}{\partial x^\mu} \; + \;
         V^i \, \frac{\partial}{\partial q^i} \; + \,
         \left( \frac{\partial V^i}{\partial q^k} \; q_\mu^k \, - \,
                \frac{\partial V^\kappa}{\partial x^\mu} \;
                q_\kappa^i \, + \,
                \frac{\partial V^i}{\partial x^\mu} \right)
         \frac{\partial}{\partial q_\mu^i}~.
\end{equation}

\subsection{Duality}

The next problem to be addressed is how to define an adequate notion of
dual for~$J^1 E$. The necessary background information from the theory of
affine spaces and of affine bundles (including the definition of the affine
dual of an affine space and of the transpose of an affine map between affine
spaces) is summarized in the Appendix. Briefly, the rules state that if $A$
is an affine space of dimension $k$ over $\mathbb{R}$, its dual $A^\star$
is the space $A(A,\mathbb{R})$ of affine maps from~$A$ to~$\mathbb{R}$,
which is a vector space of dimension $\, k+1$. \linebreak
Thus the \emph{affine dual} $J^{1\>\!\star} E$ of~$J^1 E$ and
the \emph{linear dual} $\vec{J}^{\,1\;\!\ast} E$ of $\vec{J}^{\,1} E$
are obtained by taking their fiber over any point $e$ in~$E$ to be
the vector space
\begin{equation} \label{eq:ODFJB1}
 J_e^{1\>\!\star} E~=~\{ \, z_e^{}: J_e^1 E \longrightarrow \mathbb{R}~~
                         \mathrm{affine} \, \}
\end{equation}
and
\begin{equation} \label{eq:ODLJB1}
 \vec{J}_{\,e}^{\,1\;\!\ast} E~
 =~\{ \, \vect{z}_e^{}: \vec{J}_{\,e}^{\,1} E \longrightarrow \mathbb{R}~~
         \mathrm{linear} \, \}
\end{equation}
respectively. However, as mentioned before, the multiphase spaces of field
theory are defined with an additional twist, replacing the real line by
the one-dimensional space of volume forms on the base manifold $M$ at the
appropriate point. In other words, the \emph{twisted affine dual}
\begin{equation} \label{eq:TDFJB1}
 J^{1\ostar} E~
 =~J^{1\>\!\star} E \otimes \bwedge{n} T^\ast M
\end{equation}
of $J^1 E$ and the \emph{twisted linear dual}
\begin{equation} \label{eq:TDLJB1}
 \vec{J}^{\,1\oast} E~
 =~\vec{J}^{\,1\;\!\ast} E \otimes \bwedge{n} T^\ast M
\end{equation}
of $\vec{J}^{\,1} E$ are defined by taking their fiber over any point $e$
in~$E$ with base point $\, x = \pi(e)$ \linebreak in~$M$ to be the vector
space
\begin{equation} \label{eq:TDFJB2}
 J_e^{1\ostar} E~
 =~\{ \, z_e^{}: J_e^1 E \longrightarrow \bwedge{n} T_x^* M~~
         \mathrm{affine} \, \}
\end{equation}
and
\begin{equation} \label{eq:TDLJB2}
 \vec{J}_{\,e}^{\,1\oast} E~
 =~\{ \, \vect{z}_e^{}: \vec{J}_{\,e}^{\,1} E
         \longrightarrow \bwedge{n} T_x^* M~~\mathrm{linear} \, \}
\end{equation}
respectively.\addtocounter{footnote}{-1}\footnotemark~~As in the case of the
jet bundle and the linear jet bundle, all these duals admit two different
projections, namely a \emph{target projection} onto~$E$ and a \emph{source
projection} onto~$M$ which is simply its composition with the original
projection $\pi$.

Using local coordinates as before, it is easily shown that all these duals
are fiber bundles over~$M$ with respect to the source projection (in general
without any additional structure) and are vector bundles over~$E$ with respect
to the target projection. Namely, introducing local coordinates $(x^\mu,q^i)$
for~$E$ together with the induced local coordinates $(x^\mu,q^i,q_\mu^i)$ 
for~$J^1 E$ and $(x^\mu,q^i,\vect{q}_\mu^{\,i})$ for~$\vec{J}^{\,1} E$ as
before, we obtain local coordinates $(x^\mu,q^i,p\>\!_i^\mu,p)$ both for
$J^{1\>\!\star} E$ and for $J^{1\ostar} E$ as well as local coordinates
$(x^\mu,q^i,p\>\!_i^\mu)$ both for $\vec{J}^{\,1\;\!\ast} E$ and for
$\vec{J}^{\,1\oast} E$, respectively. These are defined by requiring the
dual pairing between a point in~$J^{1\>\!\star} E$ or~$J^{1\ostar} E$
with coordinates $(x^\mu,q^i,p\>\!_i^\mu,p\>\!)$ and a point in~$J^1 E$
with coordinates $(x^\mu,q^i,q_\mu^i)$ to be given by
\begin{equation} \label{eq:POFJB}
 p\;\!_i^\mu q_\mu^i + \, p
\end{equation}
in the ordinary (untwisted) case and by
\begin{equation} \label{eq:PTFJB}
 \left( p\;\!_i^\mu q_\mu^i + \, p\;\! \right) \, d^{\,n} x
\end{equation}
in the twisted case, whereas the dual pairing between a point
in~$\vec{J}^{\,1\;\!\ast} E$ or in~$\vec{J}^{\,1\oast} E$ with
coordinates $(x^\mu,q^i,p\>\!_i^\mu)$ and a point in~$\vec{J}^{\,1} E$
with coordinates $(x^\mu,q^i,\vect{q}_\mu^{\,i})$ is given by
\begin{equation} \label{eq:POLJB}
 p\;\!_i^\mu \vect{q}_\mu^{\,i}
\end{equation}
in the ordinary (untwisted) case and by
\begin{equation} \label{eq:PTLJB}
 p\;\!_i^\mu \vect{q}_\mu^{\,i} \; d^{\,n} x
\end{equation}
in the twisted case. Moreover, a local coordinate transformation
$\, (x^\mu,q^i) \rightarrow (x^{\prime\nu},q^{\prime j})$ for~$E$
as in eq.~(\ref{eq:LCTR1}) induces local coordinate transformations
for~$J^1 E$ and for $\vec{J}^{\,1} E$ as in eqs~(\ref{eq:LCTR2}) and~%
(\ref{eq:LCTR3}) which in turn induce local coordinate transformations
$\, (x^\mu,q^i,p\>\!_i^\mu,p\>\!) \rightarrow (x^{\prime\nu},q^{\prime j},
p\;\!_j^{\prime\nu},p\;\!^\prime) \,$ both for $J^{1\>\!\star} E$
and for $J^{1\ostar} E$ as well as local coordinate transformations
$\, (x^\mu,q^i,p\>\!_i^\mu) \rightarrow (x^{\prime\nu},q^{\prime j},
p\;\!_j^{\prime\nu}) \,$ both for $\vec{J}^{\,1\;\!\ast} E$ and for
$\vec{J}^{\,1\oast} E$: these are given by
\begin{equation} \label{eq:LCTR4}
 p\;\!_j^{\prime\nu}~
 =~\frac{\partial x^{\prime\nu}}{\partial x^\mu} \;
   \frac{\partial q^i}{\partial q^{\prime j}} \; p\;\!_i^\mu~~~,~~~
 p\;\!^\prime~
 =~p \; - \, \frac{\partial q^{\prime j}}{\partial x^\mu} \;
             \frac{\partial q^i}{\partial q^{\prime j}} \; p\;\!_i^\mu
\end{equation}
in the ordinary (untwisted) case and
\begin{equation} \label{eq:LCTR5}
 p\;\!_j^{\prime\nu}~
 =~\det \Bigl( \frac{\partial x}{\partial x^\prime} \Bigr) \;
   \frac{\partial x^{\prime\nu}}{\partial x^\mu} \;
   \frac{\partial q^i}{\partial q^{\prime j}} \; p\;\!_i^\mu~~~,~~~
 p\;\!^\prime~
 =~\det \Bigl( \frac{\partial x}{\partial x^\prime} \Bigr)
   \left( p \; - \, \frac{\partial q^{\prime j}}{\partial x^\mu} \;
                    \frac{\partial q^i}{\partial q^{\prime j}} \; p\;\!_i^\mu
          \right)
\end{equation}
in the twisted case.

Finally, it is worth noting that the affine duals $J^{1\>\!\star} E$ and
$J^{1\ostar} E$ of~$J^1 E$ contain line subbundles $J_c^{1\>\!\star} E$
and $J_c^{1\ostar} E$ whose fiber over any point $e$ in~$E$ with base
point $\, x = \pi(e)$ \linebreak in~$M$ consists of the constant
(rather than affine) maps from $J_e^1 E$ to $\mathbb{R}$ and to
$\bwedge{n} T_x^* M$, respectively, and the corresponding quotient
vector bundles over~$E$ can be naturally identified with the respective
linear duals $\vec{J}^{\,1\;\!\ast} E$ and $\vec{J}^{\,1\oast} E$ of~%
$\vec{J}^{\,1} E$, i.e., we have
\begin{equation} \label{eq:OQFLJB}
 J^{1\>\!\star} E / J_c^{1\>\!\star} E~\cong~\vec{J}^{\,1\;\!\ast} E
\end{equation}
and
\begin{equation} \label{eq:TQFLJB}
 J^{1\ostar} E / J_c^{1\ostar} E~\cong~\vec{J}^{\,1\oast} E
\end{equation}
respectively. This shows that, in both cases, the corresponding projection
onto the quotient amounts to ``forgetting the additional energy variable''
since it takes a point with coordinates $(x^\mu,q^i,p\>\!_i^\mu,p\>\!)$ to
the point with coordinates $(x^\mu,q^i,p\>\!_i^\mu)$; it will be \linebreak
denoted by $\eta$ 
and is easily seen to turn $J^{1\>\!\star} E$ and $J^{1\ostar} E$
into affine line bundles over $\vec{J}^{\,1\;\!\ast} E$ and over
$\vec{J}^{\,1\oast} E$, respectively.

\subsection{The Second Order Jet Bundle}

For an appropriate global formulation of the standard Euler\,-\,Lagrange
equations of field theory, which are second order partial differential
equations, it is useful to introduce the \emph{second order jet bundle}
$J^2 E$ associated with the configuration bundle~$E$ over~$M$. \linebreak
It can be defined either directly, as is usually done, or by invoking an
iterative procedure, which is the method we shall follow here. Starting
out from the first order jet bundle $J^1 E$ of~$E$, regarded as a fiber
bundle over~$M$, we consider its first order jet bundle $J^1 J^1 E$ and
define, in a first step, the \emph{semiholonomic second order jet bundle}
$\bar{J}^2 E$ of $E$ to be the subbundle of $J^1 J^1 E$ given by
\begin{equation} \label{eq:SOJB1}
 \bar{J}^2 E~=~\{ \, \kappa \in J^1 J^1 E \, : \,
                     \tau_{J^1 E}(\kappa )~=~J^1 \tau_E(\kappa ) \, \}
\end{equation}
where $\, \tau_{J^1 E} : J^1 J^1 E \rightarrow J^1 E \,$ is the target
projection of $J^1 J^1 E$ while $\, J^1 \tau_E : J^1 J^1 E \rightarrow
J^1 E \,$ is the prolongation of the target projection $\, \tau_E :
J^1 E \rightarrow E \,$ of $J^1 E$, considered as a map of fiber bundles
over~$M$. As will become clear below, $\bar{J}^2 E$ is an affine bundle
over~$E$, with difference vector bundle $\left( T^\ast M \oplus (T^\ast M
\otimes T^\ast M) \right) \otimes V\!E$. Therefore, using the construction
of the affine quotient of an affine space (bundle) by a vector subspace
(subbundle) of its difference vector space (bundle), as explained in the
Appendix, we may complete the construction by observing that since
$T^\ast M \otimes T^\ast M$ contains $\bwedge{2} T^\ast M$ as a vector
subbundle (and hence so does $T^\ast M \oplus (T^\ast M \otimes T^\ast M)$),
it is possible to define the second order jet bundle $J^2 E$ of~$E$ as the
quotient
\begin{equation} \label{eq:SOJB2}
 J^2 E~=~\bar{J}^2 E \, / \, \bwedge{2} T^\ast M \otimes V\!E~.
\end{equation}
Once again, $J^2 E$ is an affine bundle over~$E$, with difference vector bundle
\begin{equation} \label{eq:SOJB3}
 \vec{J}^{\,2} E~
 =~\left( T^\ast M \oplus \bvee{\,2} T^\ast M \right) \otimes V\!E~.
\end{equation}

These assertions can be proved by introducing local coordinates $(x^\mu,q^i)$
for~$E$ together with the induced local coordinates $(x^\mu,q^i,q_\mu^i)$ 
for~$J^1 E$ as before to first define induced local coordinates $(x^\mu,q^i,
q_\mu^i,r_\mu^i,q_{\mu\rho}^i)$ for~$J^1 J^1 E$. Simple calculations then 
show that the points of $\bar{J}^2 E$ are characterized by the condition
$\, q_\mu^i = r_\mu^i \,$ and the points of $J^2 E$ by the additional
condition $\, q_{\mu\rho}^i = q_{\rho\mu}^i$. Moreover, a local coordinate
transformation $\, (x^\mu,q^i) \rightarrow (x^{\prime\nu},q^{\prime j}) \,$
for~$E$ as in eq.~(\ref{eq:LCTR1}) induces a local coordinate transformation
for~$J^1 E$ as in eq.~(\ref{eq:LCTR2}) which in turn induces a local
coordinate transformation $\, (x^\mu,q^i,q_\mu^i,r_\mu^i,q_{\mu\rho}^i)
\rightarrow (x^{\prime\nu}, q^{\prime j},q_\nu^{\prime j},r_\nu^{\prime j},
q_{\nu\sigma}^{\prime j}) \,$ for $J^1 J^1 E$, given by eq.~(\ref{eq:LCTR2})
together with
\begin{equation} \label{eq:LCTR6}
 r^{\prime j}_\nu~=~
 \frac{\partial x^\mu}{\partial x^{\prime \nu}} \; 
 \frac{\partial q^{\prime j}}{\partial q^i} \; r^i_\mu \; + \;
 \frac{\partial x^\mu}{\partial x^{\prime \nu}} \; 
 \frac{\partial q^{\prime j}}{\partial x^\mu}~,
\end{equation}
\begin{equation} \label{eq:LCTR7}
 q^{\prime j}_{\nu\sigma}~=~
 \frac{\partial x^\rho}{\partial x^{\prime \sigma}} \; 
 \frac{\partial q^{\prime j}_\nu}{\partial q^i_\mu} \; q^i_{\mu\rho} \; + \;
 \frac{\partial x^\rho}{\partial x^{\prime \sigma}} \; 
 \frac{\partial q^{\prime j}_\nu}{\partial x^\rho}~.
\end{equation}
In particular, eqs~(\ref{eq:LCTR2}) and (\ref{eq:LCTR6}) show that
$\, q^i_\mu = r^i_\mu \,$ implies $\, q^{\prime j}_\nu = r^{\prime j}_\nu$,
as required by the global, coordinate independent nature of the definition
of $\bar{J}^2 E$ as a subbundle of $J^1 J^1 E$, while eq.~(\ref{eq:LCTR7})
can be further evaluated by differentiating eq.~(\ref{eq:LCTR2}) with
respect to $q^i_\mu$ and $x^\rho$, which leads to
\begin{eqnarray} \label{eq:LCTR8}
 q^{\prime j}_{\nu\sigma} \!\!
 &=&\!\! \frac{\partial x^\mu}{\partial x^{\prime \nu}} \;
         \frac{\partial x^\rho}{\partial x^{\prime \sigma}} \;
         \frac{\partial q^{\prime j}}{\partial q^i} \; q^i_{\mu\rho}
                                                              \nonumber \\
 & &\!\! + \left( \frac{\partial x^\mu}{\partial x^{\prime \nu}} \;
                  \frac{\partial x^\rho}{\partial x^{\prime \sigma}} \;
                  \frac{\partial^{\>\!2} q^{\prime j}}
                       {\partial x^\rho \, \partial q^i} \; - \;
                  \frac{\partial x^\kappa}{\partial x^{\prime \nu}} \;
                  \frac{\partial x^\rho}{\partial x^{\prime \sigma}} \;
                  \frac{\partial x^\mu}{\partial x^{\prime \lambda}} \;
                  \frac{\partial^{\>\!2} x^{\prime \lambda}}
                       {\partial x^\rho \, \partial x^\kappa} \;
                  \frac{\partial q^{\prime j}}{\partial q^i} \right)
         q^i_\mu                                                        \\
 & &\!\! + \left( \frac{\partial x^\mu}{\partial x^{\prime \nu}} \;
                  \frac{\partial x^\rho}{\partial x^{\prime \sigma}} \; 
                  \frac{\partial^{\>\!2} q^{\prime j}}
                       {\partial x^\rho \, \partial x^\mu} \; - \;
                  \frac{\partial x^\kappa}{\partial x^{\prime \nu}} \;
                  \frac{\partial x^\rho}{\partial x^{\prime \sigma}} \;
                  \frac{\partial x^\mu}{\partial x^{\prime \lambda}} \;
                  \frac{\partial^{\>\!2} x^{\prime \lambda}}
                       {\partial x^\rho \, \partial x^\kappa} \;
                  \frac{\partial q^{\prime j}}{\partial x^\mu} \right)~.
                                                              \nonumber
\end{eqnarray}
This is also the induced local coordinate transformation for $\bar{J}^2 E$,
whereas that for $J^2 E$ is obtained by symmetrization:
\begin{eqnarray} \label{eq:LCTR9}
 q^{\prime j}_{\nu\sigma} \!\!
 &=&\!\! \frac{\partial x^\mu}{\partial x^{\prime \nu}} \;
         \frac{\partial x^\rho}{\partial x^{\prime \sigma}} \;
         \frac{\partial q^{\prime j}}{\partial q^i} \; q^i_{\mu\rho}
                                                              \nonumber \\
 & &\!\! + \left( \frac{1}{2} \left(
                  \frac{\partial x^\mu}{\partial x^{\prime \nu}} \;
                  \frac{\partial x^\rho}{\partial x^{\prime \sigma}} \, + \,
                  \frac{\partial x^\mu}{\partial x^{\prime \sigma}} \;
                  \frac{\partial x^\rho}{\partial x^{\prime \nu}} \right)
                  \frac{\partial^{\>\!2} q^{\prime j}}
                       {\partial x^\rho \, \partial q^i} \; - \;
                  \frac{\partial x^\kappa}{\partial x^{\prime \nu}} \;
                  \frac{\partial x^\rho}{\partial x^{\prime \sigma}} \;
                  \frac{\partial x^\mu}{\partial x^{\prime \lambda}} \;
                  \frac{\partial^{\>\!2} x^{\prime \lambda}}
                       {\partial x^\rho \, \partial x^\kappa} \;
                  \frac{\partial q^{\prime j}}{\partial q^i} \right)
         q^i_\mu                                              \nonumber \\
 & &\!\! + \left( \frac{\partial x^\mu}{\partial x^{\prime \nu}} \;
                  \frac{\partial x^\rho}{\partial x^{\prime \sigma}} \; 
                  \frac{\partial^{\>\!2} q^{\prime j}}
                       {\partial x^\rho \, \partial x^\mu} \; - \;
                  \frac{\partial x^\kappa}{\partial x^{\prime \nu}} \;
                  \frac{\partial x^\rho}{\partial x^{\prime \sigma}} \;
                  \frac{\partial x^\mu}{\partial x^{\prime \lambda}} \;
                  \frac{\partial^{\>\!2} x^{\prime \lambda}}
                       {\partial x^\rho \, \partial x^\kappa} \;
                  \frac{\partial q^{\prime j}}{\partial x^\mu} \right)~.
\end{eqnarray}
Both formulas indicate that $\bar{J}^2 E$ and $J^2 E$ are indeed affine bundles
over~$E$, with difference vector bundles as stated above.

The equivalence between the definition of the second order jet bundle given
here and the traditional one is obtained observing that the iterated jet
$j^1 j^1 \varphi$ of a (local) section $\varphi$ of $E$ assume values not
only in $\bar{J}^2 E$ but even in $J^2 E$, due to the Schwarz rule.
Therefore, second order jets in the traditional sense, that is,
classes of (local) sections where the equivalence relation is the
equality between the Taylor expansion up to second order, are in
one-to-one correspondence with these iterated jets of (local) sections.
Moreover, a global section $\varphi$ of~$E$ over~$M$ naturally induces
a global section $j^2 \varphi$ of~$J^2 E$ over~$M$ such that in the local
coordinates used above 
\[
 j^2 \varphi(x)~=~(x^\mu,\varphi^i(x),\partial_\mu \varphi^i(x),
                   \partial_\mu \partial_\nu \varphi^i(x))~.
\]
where $\partial_\mu = \partial / \partial x^\mu$; this is symbolically
summarized by writing $\, j^2 \varphi = (\varphi,\partial\varphi,
\partial^{\>\!2} \varphi)$.



\subsection{The Legendre Transformation}

A Lagrangian field theory is defined by its configuration bundle $E$
over~$M$ and its Lagrangian density or simply \emph{Lagrangian}, which
in the present first order formalism is a map of fiber bundles over~$E$:
\begin{equation} \label{eq:LAGRD}
 \mathscr{L} : J^1 E~\longrightarrow~\bwedge{n} T^\ast M~.
\end{equation}
The requirement that $\mathscr{L}$ should take values in the volume forms
rather than the functions on space-time is imposed to guarantee that the
\emph{action} functional $\, S : \Gamma(E) \rightarrow \mathbb{R} \,$ given by
\begin{equation} \label{ACTN1}
 S[\varphi]~=~\int_M \mathscr{L}(\varphi,\partial\varphi) \qquad
 \mbox{for $\, \varphi \smin\, \Gamma(E)$}
\end{equation}
be well-defined and independent of the choice of additional structures, such
as a space-time metric.\footnote{Strictly speaking, the integration should be
restricted to compact subsets of space-time, which leads to an entire family
of action functionals.} Such a Lagrangian gives rise to a \emph{Legendre
transformation}, which comes in two variants: as a map
\begin{equation} \label{eq:LEGT1}
 \vect{\mathbb{F}} \mathscr{L} : J^1 E~\longrightarrow~\vec{J}^{\,1\oast} E
\end{equation}
or as a map
\begin{equation} \label{eq:LEGT2}
 \mathbb{F} \mathscr{L} : J^1 E~\longrightarrow~J^{1\ostar} E
\end{equation}
of fiber bundles over~$E$. For any point $\gamma$ in $J_e^1 E$, the latter
is defined as the usual fiber derivative of $\mathscr{L}$ at $\gamma$, which
is the linear map from $\vec{J}_{\,e}^{\,1} E$ to $\bwedge{n} T_x^\ast M$
given by
\begin{equation} \label{eq:LEGT3}
 \vect{\mathbb{F}} \mathscr{L}(\gamma) \cdot \vect{\kappa}~
 =~\frac{d}{d\lambda} \, \mathscr{L}(\gamma +\lambda \vect{\kappa})
   \Big|_{\lambda=0} \qquad
 \mbox{for $\, \vect{\kappa} \smin \vec{J}_{\,e}^{\,1} E$}~,
\end{equation}
whereas the former encodes the entire Taylor expansion, up to first order,
of $\mathscr{L}$ around~$\gamma$ along the fibers, which is the affine map
from $J_e^1 E$ to $\bwedge{n} T_x^\ast M$ given by
\begin{equation} \label{eq:LEGT4}
 \mathbb{F} \mathscr{L}(\gamma) \cdot \kappa~
 =~\mathscr{L}(\gamma) \, + \,
   \frac{d}{d\lambda} \, \mathscr{L}(\gamma +\lambda(\kappa-\gamma))
   \Big|_{\lambda=0} \qquad
 \mbox{for $\, \kappa \smin J_e^1 E$}~.
\end{equation}
Of course, $\vect{\mathbb{F}} \mathscr{L}$ is just the linear part of
$\mathbb{F} \mathscr{L}$, that is, its composition with the bundle
projection $\eta$ from extended to ordinary multiphase space:
$\vect{\mathbb{F}} \mathscr{L} = \eta \circ \mathbb{F} \mathscr{L}$.
In local coordinates as before, $\mathbb{F} \mathscr{L}$ is given by
\begin{equation} \label{eq:LEGT5}
 p\>\!_i^\mu~=~\frac{\partial L}{\partial q_\mu^i} \quad , \quad
 p~=~L \, - \, \frac{\partial L}{\partial q_\mu^i} \, q_\mu^i
\end{equation}
where $\, \mathscr{L} = L \; d^{\,n} x$. Finally, if $\mathscr{L}$ is
supposed to be \emph{hyperregular}, which by definition means that
$\vect{\mathbb{F}} \mathscr{L}$ should be a global diffeomorphism,
then one can define the De Donder\,-\,Weyl Hamiltonian $\mathscr{H}$
to be the section of $J^{1\ostar} E$ over $\vec{J}^{\,1\oast} E$ given
by
\begin{equation} \label{eq:LEGT6}
 \mathscr{H}~
 =~\mathbb{F} \mathscr{L} \circ (\vect{\mathbb{F}} \mathscr{L})^{-1}~.
\end{equation}
In local coordinates as before, this leads to
\begin{equation} \label{eq:LEGT7}
 H~=~p\>\!_i^\mu q_\mu^i \, - \, L
\end{equation}
where $\, \mathscr{L} = L \; d^{\,n} x \,$ and $\, \mathscr{H} = - \, H \;
d^{\,n} x$, as stipulated in eq.~(\ref{eq:DWHAM}).

Conversely, the covariant Hamiltonian formulation of a field theory that
can be described in terms of a configuration bundle $E$ over~$M$ is defined
by its Hamiltonian density or simply \emph{Hamiltonian}, in the spirit of
De Donder and Weyl, which in global terms is a section of extended multi%
phase space $J^{1\ostar} E$ as an affine line bundle over ordinary multi%
phase space $\vec{J}^{\,1\oast} E$:
\begin{equation} \label{eq:HAMD}
 \mathscr{H} : \vec{J}^{\,1\oast} E~\longrightarrow~J^{1\ostar} E~.
\end{equation}
Such a Hamiltonian gives rise to an \emph{inverse Legendre transformation},
which is a map
\begin{equation} \label{eq:ILGT1}
 \mathbb{F} \mathscr{H} : \vec{J}^{\,1\oast} E~\longrightarrow~J^1 E
\end{equation}
of fiber bundles over~$E$ defined as follows. For any point $\vect{z}$ in
$\vec{J}_{\,e}^{\,1\oast} E$, the usual fiber derivative of $\mathscr{H}$
at $\vect{z}$ is a linear map from $\vec{J}_{\,e}^{\,1\oast} E$ to
$J_e^{1\ostar} E$ which when composed with the projection $\eta$ from
$J_e^{1\ostar} E$ to $\vec{J}_{\,e}^{\,1\oast} E$ gives the identity on
$\vec{J}_{\,e}^{\,1\oast} E$ (since $\mathscr{H}$ is a section): such
linear maps form an affine subspace of the vector space of all linear
maps from $\vec{J}_{\,e}^{\,1\oast} E$ to $J_e^{1\ostar} E$ that can
be naturally identified with the original affine space $J_e^1 E$, as
explained in the Appendix. In local coordinates as before, $\mathbb{F}
\mathscr{H}$ is given by
\begin{equation} \label{eq:ILGT2}
 q_\mu^i~=~\frac{\partial H}{\partial p\>\!_i^\mu}
\end{equation}
where $\, \mathscr{H} = - \, H \; d^{\,n} x$. Finally, if $\mathscr{H}$
is supposed to be \emph{hyperregular}, which by definition means that
$\mathbb{F} \mathscr{H}$ should be a global diffeomorphism, then one
can define the Lagrangian $\mathscr{L}$ to be given by
\begin{equation} \label{eq:ILGT3}
 \mathscr{L}(\gamma)~
 =~\left( \mathscr{H} \circ (\mathbb{F} \mathscr{H})^{-1} \right)(\gamma)
   \cdot \gamma~.
\end{equation}
In local coordinates as before, this leads to
\begin{equation} \label{eq:ILGT4}
 L~=~p\>\!_i^\mu q_\mu^i \, - \, H
\end{equation}
where $\, \mathscr{L} = L \; d^{\,n} x \,$ and $\, \mathscr{H} = - \, H \;
d^{\,n} x$.

Thus in the hyperregular case, the two processes are inverse to each other
and allow one to pass freely between the Lagrangian and the Hamiltonian
formulation. Of~course, this is no longer true for field theories with
local symmetries, in particular gauge theories, which require additional
conceptual input.

At any rate, it has become apparent that even in the regular case, the full
power of the multiphase space approach to geometric field theory can only
be explored if one uses the ordinary and extended multiphase spaces in
conjunction.

\subsection{Canonical Forms}

The distinguished role played by the extended multiphase space is due to
the fact that it carries a naturally defined multisymplectic form $\omega$,
derived from an equally naturally defined multicanonical form $\theta$ by
exterior differentiation: it is this property that turns it into the field
theoretical analogue of the cotangent bundle of mechanics.\footnote{Note
that this statement fails if one uses the ordinary duals instead of the
twisted ones.} Global constructions are given in the literature \cite{CCI,%
Go,GIM}, so we shall content ourselves with stating that in local coordinates
$(x^\mu,q^i,p\>\!_i^\mu,p\>\!)$ as before, $\theta$ takes the form
\begin{equation} \label{eq:MCF1}
 \theta~=~p\>\!_i^\mu \; dq^i \,\smwedge\; d^{\,n} x_\mu \, + \;
          p \; d^{\,n} x~,
\end{equation}
so $\, \omega = - d \theta \,$ becomes
\begin{equation} \label{eq:MSF1}
 \omega~=~dq^i \,\smwedge\; dp\>\!_i^\mu \,\smwedge\; d^{\,n} x_\mu \, - \;
          dp \;\smwedge\; d^{\,n} x~.
\end{equation}
Given a Lagrangian $\mathscr{L}$, we can use the associated Legendre trans%
formation $\mathbb{F} \mathscr{L}$ to pull back $\theta$ and $\omega$ and
thus define the \emph{Poincar\'e-Cartan forms} $\theta_{\mathscr{L}}^{}$
and $\omega_{\mathscr{L}}^{}$ on $J^1 E$ associated with the Lagrangian
$\mathscr{L}$:
\begin{equation}
 \theta_{\mathscr{L}}^{}~=~(\mathbb{F} \mathscr{L})^\ast \theta \quad , \quad
 \omega_{\mathscr{L}}^{}~=~(\mathbb{F} \mathscr{L})^\ast \omega~.
\end{equation}
Similarly, given a Hamiltonian $\mathscr{H}$, we can use it to pull back
$\theta$ and $\omega$ and thus define the \emph{De Donder\,-\,Weyl forms}
$\theta_{\mathscr{H}}^{}$ and $\omega_{\mathscr{H}}^{}$ on
$\vec{J}^{\,1\oast} E$ associated with the Hamiltonian
$\mathscr{H}$:
\begin{equation}
 \theta_{\mathscr{H}}^{}~=~\mathscr{H}^\ast \theta \quad , \quad
 \omega_{\mathscr{H}}^{}~=~\mathscr{H}^\ast \omega~.
\end{equation}
Of course, $\omega_{\mathscr{L}}^{} = - \, d \theta_{\mathscr{L}}^{} \,$
and $\, \omega_{\mathscr{H}}^{} = - \, d \theta_{\mathscr{H}}^{} \,$;
moreover, supposing that $\, \mathscr{H} \circ \vect{\mathbb{F}} \mathscr{L}
= \, \mathbb{F} \mathscr{L}$, we~have
\begin{equation}
 \theta_{\mathscr{L}}^{}~
 =~(\vect{\mathbb{F}} \mathscr{L})^\ast \theta_{\mathscr{H}} \quad , \quad
 \omega_{\mathscr{L}}^{}~
 =~(\vect{\mathbb{F}} \mathscr{L})^\ast \omega_{\mathscr{H}}~.
\end{equation}
In local coordinates as before, eq.~(\ref{eq:MCF1}) implies that
\begin{equation} \label{eq:MCF2}
 \theta_{\mathscr{L}}^{}~
 =~\frac{\partial L}{\partial q^i_\mu} \;
   dq^i \,\smwedge\; d^{\,n} x_\mu \, + \,
   \Big( L - \frac{\partial L}{\partial q^i_\mu} \, q^i_\mu \Big) \;
   d^{\,n} x~,
\end{equation}
\begin{equation} \label{eq:MCF3}
 \theta_{\mathscr{H}}^{}~
 =~p\>\!_i^\mu \; dq^i \,\smwedge\; d^{\,n} x_\mu \, - \;
   H \; d^{\,n} x~.
\end{equation}
It is useful to note that the forms $\theta_{\mathscr{L}}^{}$ and
$\theta_{\mathscr{H}}^{}$ allow us to give a very simple definition
of the action functional: it is given by pull-back and integration
over space-time.\addtocounter{footnote}{-1}\footnotemark~~Thus in
the Lagrangian framework, the action associated with a section
$\varphi$ of~$E$ over~$M$ is obtained by taking the pull-back of
$\theta_{\mathscr{L}}^{}$ with its derivative which is a section
$(\varphi,\partial\varphi)$ of~$J^1 E$ over~$M$,
\begin{equation} \label{eq:ACTN2}
 S[\varphi\>\!]~=~\int_M (\varphi,\partial\varphi)^\ast \,
                         \theta_{\mathscr{L}}^{} \qquad
 \mbox{for $\, \varphi \smin\, \Gamma(E)$}~,
\end{equation}
whereas in the Hamiltonian framework, the action associated with a
section $(\varphi,\pi)$ of~$\vec{J}^{\,1\oast} E$ over~$M$ is simply
\begin{equation} \label{eq:ACTN3}
 S[\varphi,\pi]~=~\int_M (\varphi,\pi)^\ast \, \theta_{\mathscr{H}}^{} \qquad
 \mbox{for $\, (\varphi,\pi) \smin\, \Gamma(\vec{J}^{\,1\oast} E)$}~.
\end{equation}
In both cases, it can be shown that the stationary points of the action
are precisely the solutions of the corresponding Euler\,-\,Lagrange and
De Donder\,-\,Weyl equations, respectively. It is therefore no surprise
that these equations can be formulated globally through the vanishing of
certain (in general nonlinear) differential operators $\mathscr{E}$ and~%
$\mathscr{D}$ defined solely in terms of the forms $\omega_{\mathscr{L}}^{}$
and $\omega_{\mathscr{H}}^{}$, respectively. However, an explicit construction
in the spirit of global analysis \cite{Pal,KMS} does not seem to be readily
available, although there do exist various attempts that go a long way in
the right direction; see, e.g., \cite{MPS} for the Lagrangian case and
\cite{CCI} for the Hamiltonian case.

\subsection{Euler\,-\,Lagrange and De Donder\,-\,Weyl Operator}

\bthe
 Given a Lagrangian density as in eq.~(\ref{eq:LAGRD}) above, define the
 corresponding \textbf{Euler\,-\,Lagrange operator} to be the map
 \begin{equation} \label{eq:EULOP1}
  \mathscr{E} : J^2 E~\longrightarrow~V^\ast E \otimes \bwedge{n} T^\ast M
 \end{equation}
 of fiber bundles over~$J^1 E$ \footnote{Again, we suppress the symbols
 indicating the pull-back of bundles from $E$ or $M$ to $J^1 E$.} that
 associates to each $2$-jet $(\varphi,\partial\varphi,\partial^{\>\!2}
 \varphi)$ of (local) sections $\varphi$ of~$E$ over~$M$ and each
 vertical vector field $\,V$ on~$E$ the $n$-form on~$M$ given by
 \begin{equation} \label{eq:EULOP2}
  \mathscr{E}(\varphi,\partial\varphi,\partial^{\>\!2} \varphi) \cdot V~
  =~(\varphi,\partial\varphi)^\ast \,
    (\mathrm{i}_{J^1 V}^{} \, \omega_\mathscr{L}^{})~.
 \end{equation} 
 Then for any section $\varphi$ of $E$, $\mathscr{E}(\varphi,\partial\varphi,
 \partial^{\>\!2} \varphi)$ is the zero section if only if $\varphi$ satisfies
 the Euler\,-\,Lagrange equations associated to $\mathscr{L}$.
\ethe
\bpro
 Let $V$ be a vertical vector field on~$E$, with local coordinate expression
 \[
  V~=~V^i \, \frac{\partial}{\partial q^i}
 \]
 (cf.~eq.~(\ref{eq:PVFE})), and let $J^1 V$ be its lifting to $J^1 E$, with
 local coordinate expression
 \[
  J^1 V~=~V^i \, \frac{\partial}{\partial q^i} \; + \,
          \left( \frac{\partial V^i}{\partial q^k} \; q_\mu^k \, + \,
                 \frac{\partial V^i}{\partial x^\mu} \right)
          \frac{\partial}{\partial q_\mu^i}
 \]
 (cf.~eq.~(\ref{eq:LVFJ})). Applying the exterior derivative to
 eq.~(\ref{eq:MCF2}), contracting with $J^1 V$ and then pulling
 back with $(\varphi,\partial\varphi)$ gives, after some
 calculation,
 \begin{eqnarray*}
  (\varphi,\partial\varphi)^\ast
  (\mathrm{i}_{J^1 V}^{} \, \omega_{\mathscr{L}}^{}) \!\!
  &=&\!\! \frac{\partial^{\,2} L}{\partial x^\mu_{} \, \partial q^i_\mu}
          (\varphi,\partial\varphi) \,
          V^i(\varphi) \; d^{\,n} x \; + \;
          \frac{\partial^{\,2} L}{\partial q^j_{} \, \partial q^i_\mu}
          (\varphi,\partial\varphi) \,
          V^i(\varphi) \, \partial_\mu^{} \varphi^j \; d^{\,n} x \\
  & &\!\! \mbox{} + \,
          \frac{\partial^{\,2} L}{\partial q^j_\nu \, \partial q^i_\mu}
          (\varphi,\partial\varphi) \,
          V^i(\varphi) \, \partial_\mu^{} \partial_\nu^{} \varphi^j \;
          d^{\,n} x \; - \;
          \frac{\partial L}{\partial q^i}(\varphi,\partial\varphi) \,
          V^i(\varphi) \; d^{\,n} x \\[2mm]
  &=&\!\! \left( \partial^{}_\mu \left( \frac{\partial L}{\partial q^i_\mu}
                                        (\varphi,\partial\varphi) \right) - \;
                 \frac{\partial L}{\partial q^i}(\varphi,\partial\varphi)
                 \right)
          V^i(\varphi) \; d^{\,n} x~,
 \end{eqnarray*}
 where it is to be noted that the terms depending on the derivatives
 of $V$ have dropped out. This leads to the following explicit formula
 for $\mathscr{E}$:
 \begin{equation} \label{eq:EULOP3}
  \mathscr{E}(\varphi,\partial\varphi,\partial^{\>\!2} \varphi)~
  = \, \left( \partial_\mu \Big( \frac{\partial L}{\partial q^i_\mu}
                                 (\varphi,\partial\varphi) \Big) \, - \,
                                 \frac{\partial L}{\partial q^i}
                                 (\varphi,\partial\varphi) \right)
        dq^i \otimes d^{\, n} x~.
 \end{equation}
 In particular, it is clear that $\mathscr{E}$ depends on $\varphi$ only
 through the point values of $\varphi$ and its partial derivatives up to
 second order, which concludes the proof.
\epro

\bthe
 Given a Hamiltonian density as in eq.~(\ref{eq:HAMD}) above, define the
 corresponding \textbf{De Donder\,-\,Weyl operator} to be the map
 \begin{equation} \label{eq:DDWOP1}
  \mathscr{D} : J^1(\vec{J}^{\,1\oast} E)~\longrightarrow~
                V^\ast(\vec{J}^{\,1\oast} E) \otimes \bwedge{n} T^\ast M
 \end{equation}
 of fiber bundles over~$\vec{J}^{\,1\oast} E$ \footnote{Again, we suppress
 the symbols indicating the pull-back of bundles from $E$ or $M$ to
 $\vec{J}^{\,1\oast} E$.} that associates to each $1$-jet $(\varphi,\pi,
 \partial\varphi,\partial\pi)$ of (local) sections $(\varphi,\pi)$
 of~$\vec{J}^{\,1\oast} E$ over~$M$ and each vertical vector field
 $\,V$ on~$\vec{J}^{\,1\oast} E$ the $n$-form on~$M$ given by
 \begin{equation} \label{eq:DDWOP2}
  \mathscr{D}(\varphi,\pi,\partial\varphi,\partial\pi) \cdot V~
  =~(\varphi,\pi)^\ast \, (\mathrm{i}_V^{} \, \omega_\mathscr{H}^{})~.
 \end{equation} 
 Then for any section $(\varphi,\pi)$ of $\vec{J}^{\,1\oast} E$,
 $\mathscr{D}(\varphi,\pi,\partial\varphi,\partial\pi)$ is the
 zero section if only if $(\varphi,\pi)$ satisfies the
 De Donder\,-\,Weyl equations associated to $\mathscr{H}$.
\ethe
\bpro
 Let $V$ be a vertical vector field on~$\vec{J}^{\,1\oast} E$, with local
 coordinate expression
 \[
  V~=~V^i \, \frac{\partial}{\partial q^i} \; + \;
      V_i^\mu \, \frac{\partial}{\partial p\>\!^\mu_i}~.
 \]
 Applying the exterior derivative to eq.~(\ref{eq:MCF3}), contracting with $V$
 and then pulling back with $(\varphi,\pi)$ gives, after a short calculation,
 \begin{eqnarray*}
  (\varphi,\pi)^\ast (\mathrm{i}_V^{} \, \omega_{\mathscr{H}}^{}) \!\!
  &=&\!\! \partial_\mu \pi_i^\mu \; V^i(\varphi,\pi) \; d^{\,n} x \; + \;
          \frac{\partial H}{\partial q^i}(\varphi,\pi) \;
          V^i(\varphi,\pi) \; d^{\,n} x \\
  & &\!\! \mbox{} - \, \partial_\mu \varphi^i \;
          V_i^\mu(\varphi,\pi) \; d^{\,n} x \; + \;
          \frac{\partial H}{\partial p\>\!_i^\mu}(\varphi,\pi) \;
          V_i^\mu(\varphi,\pi) \; d^{\,n} x~.
 \end{eqnarray*}
 This leads to the following explicit formula for $\mathscr{D}$:
 \begin{equation} \label{eq:DDWOP3}
  \begin{array}{rcl}
   {\displaystyle
    \mathscr{D}(\varphi,\pi,\partial\varphi,\partial\pi) \!\!}
   &=&\!\! {\displaystyle
            \left( \frac{\partial H}{\partial q^i}(\varphi,\pi) \, - \,
                   \partial_\mu \pi_i^\mu \right)
            dq^i \otimes d^{\, n} x} \\[6mm]
   & &\!\! {\displaystyle \mbox{} +
            \left( \frac{\partial H}{\partial p\>\!_i^\mu}(\varphi,\pi) \, + \,
                   \partial_\mu \varphi^i \right)
            dp\>\!_i^\mu \otimes d^{\, n} x~.}
  \end{array}
 \end{equation}
 In particular, it is clear that $\mathscr{D}$ depends on $(\varphi,\pi)$
 only through the point values of $\varphi$ and $\pi$ and their partial
 derivatives up to first order, which concludes the proof.
\epro

\subsection{Jacobi Operators}

In order to make contact with the functional formalism to be discussed in the
next section, we must also derive explicit expressions for the linearization
of the Euler\,-\,Lagrange operator and the De Donder\,-\,Weyl operator around
a given solution of the equations of motion. This leads to \emph{linear}
differential operators between vector bundles over~$M$ that we shall
refer to as \emph{Jacobi operators}, generalizing the familiar derivation
of the Jacobi equation by linearizing the geodesic equation.

In its Lagrangian version, the \emph{Jacobi operator} is a second order
differential operator
\begin{equation}
 \mathscr{J}_{\mathscr{L}}^{}[\varphi\>\!] :
 \Gamma(V_\varphi)~\longrightarrow~\Gamma(V_\varphi^{\oast})~,
\end{equation}
where $\, V_\varphi = \varphi^\ast (V\!E) \,$ and $\, V_\varphi^{\oast} =
\varphi^\ast (V^\ast E) \otimes \bwedge{n} T^\ast M$, obtained by linearizing
the Euler\,-\,Lagrange operator $\mathscr{E}$  around a given solution
$\varphi$ of the equations of motion. \linebreak Similarly, in its
Hamiltonian version, the \emph{Jacobi operator} is a first order
differential operator
\begin{equation}
 \mathscr{J}_{\mathscr{H}}^{}[\varphi,\pi] :
 \Gamma(V_{(\varphi,\pi)})~\longrightarrow~\Gamma(V_{(\varphi,\pi)}^{\oast})~,
\end{equation}
where $\, V_{(\varphi,\pi)} = (\varphi,\pi)^\ast (V(\vec{J}^{\,1\oast} E)) \,$
and $\, V_{(\varphi,\pi)}^{\oast} = (\varphi,\pi)^\ast (V^\ast(\vec{J}^%
{\,1\oast} E)) \otimes \bwedge{n} T^\ast M$, obtained by linearizing the
De Donder\,-\,Weyl operator $\mathscr{D}$  around a given solution
$(\varphi,\pi)$ of the equations of motion. (Thus in both cases, the
vector bundles involved are obtained by pulling back the appropriate
vertical bundle and its twisted dual with the solution of the nonlinear
equation around which the linearization is performed,) To obtain explicit
expressions, consider an arbitrary variation $\varphi_\lambda$ around
$\varphi$ and evaluate $\mathscr{E}(\varphi_\lambda,\partial\varphi_%
\lambda,\partial^{\>\!2} \varphi_\lambda)$ which, for each $\lambda$,
is a section of $V_{\varphi_\lambda}^{\oast}$, observing that since
$\, \varphi = \varphi_\lambda \big|_{\lambda=0} \,$ is a solution,
$\mathscr{E}(\varphi_\lambda,\partial\varphi_\lambda,\partial^{\>\!2}
\varphi_\lambda) \big|_{\lambda=0}$ is the zero section of
$V_\varphi^{\oast}$, and setting
\begin{equation} \label{eq:LVAR1}
 \delta\varphi~=~\frac{\partial}{\partial\lambda} \, \varphi_\lambda \,
                 \Big|_{\lambda=0}~.
\end{equation}
Noting that in local coordinates, the value of $\mathscr{E}(\varphi_%
\lambda,\partial\varphi_\lambda,\partial^{\>\!2} \varphi_\lambda)$ at
a point $x$ in~$M$ with coordinates $x^\mu$ has coordinates $(x^\mu,
\varphi_\lambda^i(x),\mathscr{E}(\varphi_\lambda,\partial\varphi_%
\lambda,\partial^{\>\!2} \varphi_\lambda)_i(x))$ where the last
piece is the coefficient of $\, dq^i \otimes d^{\, n} x \,$ in
eq.~(\ref{eq:EULOP3}), we get by differentiation with respect
to $\lambda$
\begin{eqnarray*}
\lefteqn{\frac{\partial}{\partial \lambda} \,
         \mathscr{E}(\varphi_\lambda,\partial\varphi_\lambda,
                     \partial^{\>\!2} \varphi_\lambda) \Big|_{\lambda=0}}
                                                           \hspace{5mm} \\[2mm]
 &=&     \delta \varphi^i \, \frac{\partial}{\partial q^i}              \\[2mm]
 & &\!\! \mbox{} + \,
         \bigg( \partial_\mu \Big( \frac{\partial^{\>\!2} L}
                                        {\partial q^j \, \partial q^i_\mu}
                                   (\varphi,\partial\varphi) \;
                                   \delta \varphi^j \, + \,
                                   \frac{\partial^{\>\!2} L}
                                        {\partial q^j_\nu \, \partial q^i_\mu}
                                   (\varphi,\partial\varphi) \;
                                   \partial_\nu \delta \varphi^j \Big)  \\
 & &\!\! \hspace{10mm} - \,
                \frac{\partial^{\>\!2} L}{\partial q^j \, \partial q^i}
                (\varphi,\partial\varphi) \; \delta \varphi^j \, - \,
                \frac{\partial^{\>\!2} L}{\partial q_\nu^j \, \partial q^i}
                (\varphi,\partial\varphi) \; \partial_\nu \delta \varphi^j
         \bigg) \, dq^i \otimes d^{\, n} x~.
\end{eqnarray*}
Similarly, consider an arbitrary variation $(\varphi_\lambda,\pi_\lambda)$
around $(\varphi,\pi)$ and evaluate $\mathscr{D}(\varphi_\lambda,\pi_\lambda,
\partial\varphi_\lambda,\partial\pi_\lambda)$ which, for each $\lambda$, is a
section of $V_{(\varphi_\lambda,\pi_\lambda)}^{\oast}$, observing that since
$\, (\varphi,\pi) = (\varphi_\lambda,\pi_\lambda) \big|_{\lambda=0} \,$ is
a solution, $\mathscr{D}(\varphi_\lambda,\pi_\lambda,\partial\varphi_%
\lambda,\partial\pi_\lambda) \big|_{\lambda=0}$ is the zero section of
$V_{(\varphi,\pi)}^{\oast}$, and setting
\begin{equation} \label{eq:HVAR1}
 (\delta\varphi,\delta\pi)~
 =~\frac{\partial}{\partial\lambda} \, (\varphi_\lambda,\pi_\lambda) \,
   \Big|_{\lambda=0}~.
\end{equation}
Again, noting that in local coordinates, the value of $\mathscr{D}(\varphi_%
\lambda,\pi_\lambda,\partial\varphi_\lambda,\partial\pi_\lambda)$ at
a point $x$ in~$M$ with coordinates $x^\mu$ has coordinates $(x^\mu,
\varphi_\lambda^i(x),(\pi_\lambda)_i^\mu(x),\mathscr{D}(\varphi_\lambda,
\pi_\lambda,\partial\varphi_\lambda,\partial\pi_\lambda)_i(x), \linebreak
\mathscr{D}(\varphi_\lambda,\pi_\lambda,\partial\varphi_\lambda,
\partial\pi_\lambda)_\mu^i(x))$ where the last two pieces are the
coefficients of $\, dq^i \otimes d^{\, n} x \,$ and of $\, dp\>\!_i^\mu
\otimes d^{\, n} x \,$ in eq.~(\ref{eq:DDWOP3}), we get by differentiation
with respect to $\lambda$
\begin{eqnarray*}
\lefteqn{\frac{\partial}{\partial \lambda} \,
         \mathscr{D}(\varphi_\lambda,\pi_\lambda,
                     \partial\varphi_\lambda,\partial\pi_\lambda)
         \Big|_{\lambda=0}}                                \hspace{5mm} \\[2mm]
 &=&     \delta \varphi^i \, \frac{\partial}{\partial q^i} \, + \,
         \delta \pi_i^\mu \, \frac{\partial}{\partial p\>\!_i^\mu}      \\[2mm]
 & &\!\! \mbox{} + \left( \frac{\partial^{\>\!2} H}
                               {\partial q^j \, \partial q^i}
                          (\varphi,\pi) \; \delta \varphi^j \, + \,
                          \frac{\partial^{\>\!2} H}
                               {\partial p\>\!_j^\nu \, \partial q^i}
                          (\varphi,\pi) \; \delta \pi_j^\nu \, - \,
                          \partial_\mu \, \delta \pi_i^\mu \right)
         dq^i \otimes \, d^{\, n} x \\
 & &\!\! \mbox{} + \left( \frac{\partial^{\>\!2} H}
                               {\partial q^j \, \partial p\>\!_i^\mu}
                          (\varphi,\pi) \; \delta \varphi^j \, + \,
                          \frac{\partial^{\>\!2} H}
                               {\partial p\>\!_j^\nu \,
                                \partial p\>\!_i^\mu} \, (\varphi,\pi) \;
                          \delta \pi_j^\nu \, + \,
                          \partial_\mu \, \delta \varphi^i \right)
         dp\>\!_i^\mu \otimes d^{\, n} x~.
\end{eqnarray*}

In order to show how to extract the Jacobi operators from these expressions, by
means of a globally defined prescription, we apply the following construction~%
\cite{GoS}. \linebreak
Let $F$ be a fiber bundle over~$M$, with bundle projection $\, \pi_{F,M} : F
\rightarrow M$, and $W$ be a vector bundle over~$F$ with bundle projection
$\, \pi_{W,F} : W \rightarrow F$, which is then also a fiber bundle (but not
necessarily a vector bundle) over~$M$ with respect to the composite bundle
projection $\, \pi_{W,M} = \pi_{F,M} \circ \pi_{W,F} : W \rightarrow M$.
Thus $W$ admits two different kinds of vertical bundles, $V_F W$ and $V_M W$,
with fibers defined by $\, (V_F)_w W = \ker \, T_w \pi_{W,F} \,$ and
$\, (V_M)_w W = \ker \, T_w \pi_{W,M} \,$ for $\, w \smin W$; obviously,
the former is contained in the latter as a vector subbundle. Moreover,
since $W$ is supposed to be a vector bundle over~$F$, there is a canonical
isomorphism $\, V_F W \cong \pi_{W,F}^\ast W$. On the other hand, consider
the vertical bundle $VF$ of~$F$ which can be pulled back to~$W$ to obtain
a vector bundle $\pi_{W,F}^\ast(VF)$ over~$W$, with fibers defined by
$\, (\pi_{W,F}^\ast(VF))_w = V_f F = \ker \, T_f \pi_{F,M} \,$ for
$\, w \smin W \,$ with $\, f = \pi_{W,F} w$. Note also that the tangent
map to the bundle projection $\pi_{W,F}$, which by definition has kernel
$V_F W$, maps $V_M W$ onto $VF$, so we have the following exact sequence
of vector bundles over~$W$:
\[
 0~\longrightarrow~V_F W \cong \pi_{W,F}^\ast W~\longrightarrow~V_M W~
   \longrightarrow~\pi_{W,F}^\ast (VF)~\longrightarrow~0~.
\]
The crucial observation is now that this exact sequence admits a canonical
splitting over the zero section $\, 0 : F \rightarrow W$, given simply by
its tangent map. Indeed, its tangent map $\, T_f 0 : T_f F \rightarrow
T_{0(f)} W \,$ at any point $\, f \smin F \,$ takes the vertical subspace
$V_f F$ to the $M$-vertical subspace $(V_M)_{0(f)} W$ and so restricts to a
vertical tangent map $\, V_f 0 : V_f F \rightarrow (V_M)_{0(f)} W \,$ whose
composition with the restriction of the tangent map $\, T_{0(f)} \pi_{W,F} :
T_{0(f)} W \rightarrow T_f F \,$ to $(V_M)_{0(f)} W$ gives the identity on
$V_f F$. Thus the image of $V_f 0$ is a subspace of $(V_M)_{0(f)} W$ that
is complementary to the subspace $\, (V_F)_{0(f)} W = W_f \,$ and provides
a surjective linear map $\, \sigma_f : (V_M)_{0(f)} W \rightarrow W_f \,$
of which it is the kernel. At the level of bundles, this corresponds to
a surjective vector bundle homomorphism $\, \sigma : V_M W \big|_{0}
\rightarrow W$.

Applying this construction to the situation at hand, take $\, F = E \,$ in
the Lagrangian case and $\, F = \vec{J}^{\,1\oast} E \,$ in the Hamiltonian
case, setting $\, W = V^\ast(F) \otimes \bwedge{n} T^\ast M \,$ in both cases.
The fact that the Euler\,-\,Lagrange or De Donder\,-\,Weyl operator is being
linearized around a solution $\varphi$ or $(\varphi,\pi)$ of the equations
of motion then means that we are evaluating its derivative, which a priori
takes the variation $\delta\varphi$ or $(\delta\varphi,\delta\pi)$ to a
vector field on $W$ along~$M$ which is vertical with respect to the
projection of $W$ onto~$M$, precisely over the zero section, so we
can apply the operator $\sigma$ just introduced to project it down
to a section of $W$ over~$M$ itself. This operation completes the
definition of the Jacobi operators, namely
\begin{equation} \label{eq:LJOP1}
 \mathscr{J}_{\mathscr{L}}^{}[\varphi\>\!] \cdot \delta\varphi~
 =~\sigma \left( \frac{\partial}{\partial\lambda} \,
                 \mathscr{E}(\varphi_\lambda,\partial\varphi_\lambda,
                             \partial^{\>\!2} \varphi_\lambda)
                 \Big|_{\lambda=0} \right) \!~,
\end{equation}
and
\begin{equation} \label{eq:HJOP1}
 \mathscr{J}_{\mathscr{H}}^{}[\varphi,\pi] \cdot (\delta\varphi,\delta\pi)~
 =~\sigma \left( \frac{\partial}{\partial\lambda} \,
                 \mathscr{D}(\varphi_\lambda,\pi_\lambda,
                             \partial\varphi_\lambda,\partial\pi_\lambda)
                 \Big|_{\lambda=0} \right) \!~.
\end{equation}
The local coordinate expressions are the ones derived above, that is,
\begin{eqnarray} \label{eq:LJOP2}
 \mathscr{J}_{\mathscr{L}}^{}[\varphi\>\!] \cdot \delta\varphi \!\!
 &=&\!\! \bigg\{ \frac{\partial^{\>\!2} L}
                      {\partial q^j_\nu \, \partial q^i_\mu}
                      (\varphi,\partial\varphi) \;
                 \partial_\mu \partial_\nu \delta \varphi^j   \nonumber \\[2mm]
 & &\!\! \hspace{5mm} + \,
         \bigg( \partial_\mu \Big( \frac{\partial^{\>\!2} L}
                                        {\partial q^j_\nu \, \partial q^i_\mu}
                                   (\varphi,\partial\varphi) \Big) \, + \,
                             \Big( \frac{\partial^{\>\!2} L}
                                        {\partial q^j \,
                                         \partial q^i_\nu} \, - \,
                                   \frac{\partial^{\>\!2} L}
                                        {\partial q^i \,
                                         \partial q^j_\nu} \Big)
                                   (\varphi,\partial\varphi) \bigg) \,
         \partial_\nu \delta \varphi^j \qquad                           \\[2mm]
 & &\!\! \hspace{5mm} + \,
         \bigg( \partial_\mu \Big( \frac{\partial^{\>\!2} L}
                                        {\partial q^j \, \partial q_\mu^i}
                                   (\varphi,\partial\varphi) \Big) \, - \,
                                   \frac{\partial^{\>\!2} L}
                                        {\partial q^j \, \partial q^i}
                                   (\varphi,\partial\varphi) \bigg) \,
         \delta \varphi^j \, \bigg\} \; dq^i \otimes d^{\,n} x~, \nonumber
\end{eqnarray}
and
\begin{eqnarray} \label{eq:HJOP2}
\lefteqn{\mathscr{J}_{\mathscr{H}}^{}[\varphi,\pi] \cdot
 (\delta\varphi,\delta\pi)}                      \hspace{5mm} \nonumber \\[4mm]
 &=&\!\! \left( \frac{\partial^{\>\!2} H}
                     {\partial q^j \, \partial q^i}
                (\varphi,\pi) \; \delta \varphi^j \, + \,
                \frac{\partial^{\>\!2} H}
                     {\partial p\>\!_j^\nu \, \partial q^i}
                (\varphi,\pi) \; \delta \pi_j^\nu \, - \,
                \partial_\mu \, \delta \pi_i^\mu \right)
         dq^i \otimes \, d^{\, n} x                                     \\[2mm]
 & &\!\! \mbox{} +
         \left( \frac{\partial^{\>\!2} H}
                     {\partial q^j \, \partial p\>\!_i^\mu}
                (\varphi,\pi) \; \delta \varphi^j \, + \,
                \frac{\partial^{\>\!2} H}
                     {\partial p\>\!_j^\nu \, \partial p\>\!_i^\mu} \,
                (\varphi,\pi) \; \delta \pi_j^\nu \, + \,
                \partial_\mu \, \delta \varphi^i \right)
         dp\>\!_i^\mu \otimes d^{\,n} x~.                     \nonumber
\end{eqnarray}

\section{Functional Approach}

Let us begin by recalling the definition of the Poisson bracket between
functions on a symplectic manifold with symplectic form $\omega$. First, 
one associates to each (smooth) function $f$ a (smooth) Hamiltonian
vector field $X_f$, uniquely determined by the condition
\begin{equation} \label{eq:HAMVF1}
 \mathrm{i}_{X_f}\omega~=~df~.
\end{equation}
Then the Poisson bracket of two functions $f$ and $g$ is defined to
be the function $\{f,g\}$ given by
\begin{equation} \label{eq:POISB1}
 \{f,g\}~=~- \mathrm{i}_{X_f} \mathrm{i}_{X_g} \omega~
 =~\mathrm{d}f(X_g)~= \; - \, \mathrm{d}g(X_f)~.
\end{equation}
The goal of this section is to show that formally, the same construction
applied to covariant phase space links the Witten symplectic form to the
Peierls bracket.

\subsection{Covariant Phase Space}

In contrast to the traditional non-covariant Hamiltonian formalism of field
theory, where phase space is a ``space'' of Cauchy data, \emph{covariant phase
space}, denoted here by $\mathscr{S}$, is the ``space'' of solutions of the
equations of motion, or field equations. Of course, one cannot expect these
two interpretations of phase space to be equivalent in complete generality,
since it is well known that, for nonlinear equations, time evolution of
regular Cauchy data may lead to solutions that, within finite time, develop
some kind of singularity. An even more elementary prerequisite is that the
underlying space-time manifold $M$ must admit at least some Cauchy surface
$\Sigma\,$: this means that $M$ should be globally hyperbolic.

Thus our basic assumption for the remainder of this paper will be that the
underlying space-time manifold $M$ should be \emph{globally hyperbolic}.
Globally hyperbolic space-times are the natural arena for the mathematical
theory of hyperbolic (systems of) partial differential equations, in which
the Cauchy problem is well posed. There are by now various and apparently
quite different definitions of the concept of a globally hyperbolic space-%
time, but they have ultimately turned out to be all equivalent; see 
Chapter~8 of \cite{Wa} for an extensive discussion. For our purposes,
the most convenient one is that~$M$ admits a global time function whose
level surfaces provide a foliation of $M$ into Cauchy surfaces, providing
a global diffeomorphism $\, M \cong \mathbb{R} \times \Sigma$. As an
immediate corollary, we can define the concept of a (closed/open)
\emph{time slice} in $M$: it is a (closed/open) subset of $M$ which
under such a global diffeomorphism corresponds to a subset of the
form $\, I \times \Sigma \,$ where $I$ is a (closed/open) interval
in $\mathbb{R}$.

In the Lagrangian as well as in the Hamiltonian approach to field theory,
the equations of motion are derived from a variational principle, that is,
their solutions are the stationary points of a certain functional $S$
called the \emph{action} and defined on a space of sections of an
appropriate fiber bundle over space-time which is usually referred to
as the \emph{space of field configurations} of the theory and will in what
follows be denoted by~$\mathscr{C}$. More concretely, $\mathscr{C}$ is the
space $\Gamma(F)$ of smooth sections $\phi$ of a fiber bundle $F$ over~$M$:
in the Lagrangian approach, $F$ is the configuration bundle $E$, whereas in
the Hamiltonian approach, $F$ is the multiphase space $\vec{J}^{\,1\,\oast} E$,
regarded as a fiber bundle over~$M$.

Formally, we shall as usual think of $\mathscr{C}$ as being a manifold
(which is of course infinite-dimensional). As such, it has at each of its
points $\phi$ a tangent space $T_\phi \mathscr{C}$ that can be defined
formally as a space of smooth sections, with appropriate support
properties, of the vector bundle $\, V_\phi = \phi^\ast (V\!F) \,$
over~$M$, i.e., $T_\phi \mathscr{C}~\smsubset~\Gamma^\infty(V_\phi)$.
The cotangent space $T_\phi^\ast \mathscr{C}$ will then be the space of 
distributional sections, with dual support properties, of the vector bundle
$\, V_\phi^{\oast} = \phi^\ast (V^\ast F) \otimes \bwedge{n} T^\ast M \,$
over~$M$, i.e., $T_\phi^\ast \mathscr{C}~\smsubset~\Gamma^{-\infty}
(V_\phi^{\oast\,})$. It contains as a subspace the corresponding space
of smooth sections, where the pairing between a smooth section of
$V_\phi$ and a smooth section of $V_\phi^{\oast}$ (with appropriate
support conditions) is given by contraction and integration of the
resulting form over~$M$. Similarly, the second tensor power
$\, T_\phi^\ast \mathscr{C} \otimes T_\phi^\ast \mathscr{C}$
of $T_\phi^\ast \mathscr{C}$ can be thought of as the space of
distributional sections, again with dual support properties, of
the second exterior tensor power%
\footnote{If $V$ and $W$ are vector bundles over $M$, $V \boxtimes W \,$
is defined to be the vector bundle over~$\, M \times M$ with fibers given
by $\, (V \!\boxtimes W)_{(x,y)} = V_x \otimes W_y$, for all $\, x,y \smin M$.}
\linebreak
$\, V_\phi^{\oast} \;\!\boxtimes V_\phi^{\oast} \,$ of $V_\phi^{\oast}$\,;
it contains as a subspace the corresponding space of smooth sections,
where the pairing between a pair of smooth sections of $V_\phi$ and a
smooth section of $\, V_\phi^{\oast} \;\!\boxtimes V_\phi^{\oast}$
(with appropriate support conditions) is given by contraction and
integration of the resulting form over~$\, M \times M$.

Regarding the support conditions to be imposed, the first two options
that come to mind would be to require that either the elements of $T_\phi
\mathscr{C}$ or the elements of $T_\phi^\ast \mathscr{C}$ should have
compact support, which would imply that the support of the elements of
the corresponding dual, $T_\phi^\ast \mathscr{C}$ or $T_\phi \mathscr{C}$,
could be left completely arbitrary:
\begin{equation} \label{eq:TCS1}
 \textsc{Option 1}: \qquad \qquad
 T_\phi \mathscr{C}~=~\Gamma^\infty(V_\phi) \quad , \quad
 T_\phi^\ast \mathscr{C}~=~\Gamma_c^{-\infty}(V_\phi^{\oast\,}) \qquad
\end{equation}
\begin{equation} \label{eq:TCS2}
 \textsc{Option 2}: \qquad \qquad
 T_\phi \mathscr{C}~=~\Gamma_c^\infty(V_\phi) \quad , \quad
 T_\phi^\ast \mathscr{C}~=~\Gamma^{-\infty}(V_\phi^{\oast\,}) \qquad
\end{equation}
There is a third option that makes use of the assumption that $M$ is  globally
hyperbolic. To formulate it, we introduce the following terminology. A section
of a vector bundle over~$M$ is said to have \emph{spatially compact support}
if the intersection between its support and any (closed) time slice in~$M$
is compact, and it is said to have \emph{temporally compact support} if its
support is contained in some time slice. Then, as in Ref.~\cite{KS1}, we
require the elements of $T_\phi \mathscr{C}$ to have spatially compact
support and the elements of $T_\phi^\ast \mathscr{C}$ to have temporally
compact support:
\begin{equation} \label{eq:TCS3}
 \textsc{Option 3}: \qquad \qquad
 T_\phi \mathscr{C}~=~\Gamma_{sc}^\infty(V_\phi) \quad , \quad
 T_\phi^\ast \mathscr{C}~=~\Gamma_{tc}^{-\infty}(V_\phi^{\oast\,}) \qquad
\end{equation}
Obviously, for each of these three options, the two spaces listed above
are naturally dual to each other.%
\footnote{Here and in what follows, the symbols $\Gamma_c$, $\Gamma_{sc}$ and
$\Gamma_{tc}$ indicate spaces of sections of compact, spatially compact and
temporally compact support, respectively.}

These constructions can be applied to elucidate the nature of functional
derivatives of functionals on~$\mathscr{C}$, such as the action. Namely,
given a (formally smooth) functional $\, \mathslf{F} : \mathscr{C} \rightarrow
\mathbb{R}$, its functional derivative at a point $\phi$ is the linear
functional on~$T_\phi \mathscr{C}$ which, when applied to $\delta\phi\,$,
yields the directional derivative of $\mathslf{F}$ at $\phi$ along
$\delta\phi\,$, defined by the requirement that for any one-parameter
family of sections $\phi_\lambda$ of $F$ such that $\, \phi_\lambda
\big|_{\lambda=0} = \phi\,$,
\[
 \mathslf{F}^{\,\prime}[\phi] \cdot \delta\phi~
 =~\frac{d}{d\lambda} \, \mathslf{F}\,[\phi_\lambda] \, \Big|_{\lambda=0}
 \qquad \mbox{if} \qquad
 \delta\phi~=~\frac{\partial}{\partial\lambda} \, \phi_\lambda \,
              \Big|_{\lambda=0}~.
\]
Then $\mathslf{F}^{\,\prime}[\phi]$ is a distributional section of
$V_\phi^{\oast}$ with appropriate support properties (dual to those
required for $T_\phi \mathscr{C}$). In local coordinates, its action
on $\delta\phi$ can (formally and at least when the intersection of
the two supports is contained in the coordinate system domain) be
written in the form
\begin{equation} \label{eq:FUNDER1}
 \mathslf{F}^{\,\prime}[\phi] \cdot \delta\phi~
 =~\int_M d^{\, n} x~\frac{\delta \mathslf{F}}{\delta\phi}[\phi](x) \cdot
                     \delta\phi(x)~,
\end{equation}
The expression $(\delta \mathslf{F} / \delta\phi)[\phi]$, sometimes called the
variational derivative of $\mathslf{F}$ at $\phi\,$, is then a distributional
section of~$V_\phi^\ast$ (over the coordinate system domain). In the
Lagrangian framework,
\[
 \frac{\delta \mathslf{F}}{\delta\phi}[\phi](x)~
 =~\frac{\delta \mathslf{F}}{\delta\varphi^i}[\varphi](x)~dq^i~,
\]
whereas in the Hamiltonian framework,
\[
 \frac{\delta \mathslf{F}}{\delta\phi}[\phi](x)~
 =~\frac{\delta \mathslf{F}}{\delta\varphi^i}[\varphi,\pi](x)~dq^i \, + \,
   \frac{\delta \mathslf{F}}{\delta\pi_i^\mu}[\varphi,\pi](x)~dp\>\!_i^\mu~.
\]
Similarly, the second functional derivative of $\mathslf{F}$ at
$\phi$ is the symmetric bilinear functional on $T_\phi \mathscr{C}$
which, when applied to $\delta\phi_1^{}$ and $\delta\phi_2^{}$,
can be defined by the requirement that for any two-parameter
family of sections $\phi_{\lambda_1,\lambda_2}$ of $F$ such that
$\, \phi_{\lambda_1,\lambda_2} \big|_{\lambda_1,\lambda_2=0} = \phi\,$,
\[
 \mathslf{F}^{\,\prime\prime}[\phi] \cdot (\delta\phi_1^{},\delta\phi_2^{})~
 =~\frac{\partial^{\>\!2}}{\partial\lambda_1 \, \partial\lambda_2} \,
   \mathslf{F}\,[\phi_{\lambda_1,\lambda_2}] \, \Big|_{\lambda_1,\lambda_2=0}
\]
if
\[
 \delta\phi_1^{}~=~\frac{\partial}{\partial\lambda_1} \,
 \phi_{\lambda_1,\lambda_2} \, \Big|_{\lambda_1,\lambda_2=0}~~,~~
 \delta\phi_2^{}~=~\frac{\partial}{\partial\lambda_2} \,
 \phi_{\lambda_1,\lambda_2} \, \Big|_{\lambda_1,\lambda_2=0}~.
\]
Then $\mathslf{F}^{\,\prime\prime}[\phi]$ is a distributional section of
$\, V_\phi^{\oast} \;\!\boxtimes V_\phi^{\oast} \,$ with appropriate support
properties (dual to those required for $\, T_\phi \mathscr{C} \otimes
T_\phi \mathscr{C}$). In local coordinates for $\, M \times M \,$
induced from local coordinates for $M$, its action on $(\delta\phi_1^{},
\delta\phi_2^{})$ can (formally and at least when the intersection of the
supports is contained in the coordinate system domain) be written in the form
\begin{equation} \label{eq:FUNDER2}
 \mathslf{F}^{\,\prime\prime}[\phi] \cdot (\delta\phi_1^{},\delta\phi_2^{})~
 =~\int_M d^{\, n} x \int_M d^{\, n} y~
   \frac{\delta^2 \mathslf{F}}{\delta\phi^2}[\phi](x,y) \cdot 
   (\delta\phi_1^{}(x),\delta\phi_2^{}(y))~, 
\end{equation}
The expression $(\delta^2 \mathslf{F} / \delta\phi^2)[\phi]$, sometimes called
the variational Hessian of $\mathslf{F}$ at $\phi\,$, is then a distributional
section of~$\, V_\phi^\ast \boxtimes V_\phi^\ast \,$ (over the coordinate
system domain). In the Lagrangian framework,
\[
 \frac{\delta^2 \mathslf{F}}{\delta\phi^2}[\phi](x,y)~
 =~\frac{\delta^2 \mathslf{F}}{\delta\varphi^i \, \delta\varphi^j}[\varphi](x,y)~
   dq^i \otimes dq^j~,
\]
whereas in the Hamiltonian framework,
\begin{eqnarray*}
 \frac{\delta^2 \mathslf{F}}{\delta\phi^2}[\phi](x,y) \!\!
 &=&\!\! \frac{\delta^2 \mathslf{F}}{\delta\varphi^i \, \delta\varphi^j}
         [\varphi,\pi](x,y)~dq^i \otimes dq^j~+~
         \frac{\delta^2 \mathslf{F}}{\delta\varphi^i \, \delta\pi_j^\nu}
         [\varphi,\pi](x,y)~dq^i \otimes dp\>\!_j^\nu \\
 & &\!\! \mbox{} + \,
         \frac{\delta^2 \mathslf{F}}{\delta\pi_i^\mu \, \delta\varphi^j}
         [\varphi,\pi](x,y)~dp\>\!_i^\mu \otimes dq^j~+~
         \frac{\delta^2 \mathslf{F}}{\delta\pi_i^\mu \, \delta\pi_j^\nu}
         [\varphi,\pi](x,y)~dp\>\!_i^\mu \otimes dp\>\!_j^\nu~.
\end{eqnarray*}

Of course, for the integrals in eqs~(\ref{eq:FUNDER1}) and~(\ref{eq:FUNDER2})
to make sense, even when interpreted in the sense of pairing distributions
with test functions, we must make some assumption about support properties,
which leads us back to the options stated in eqs~(\ref{eq:TCS1})-%
(\ref{eq:TCS3}). \linebreak
Option~1: when $\mathslf{F}\,$ is \emph{arbitrary}, we have to restrict the
sections $\delta\phi$, $\delta\phi_1^{}$, $\delta\phi_2^{}$ of $V_\phi$
considered above to have compact support (which can be achieved if the
sections $\phi_\lambda$, $\phi_{\lambda_1,\lambda_2}$ of $F$ are supposed
to be independent of the parameters outside a compact subset). \linebreak
Option~2: when $\mathslf{F}\,$ is \emph{local}, which we understand to mean
that its functional dependence on the fields is non-trivial only within a
compact region, or equivalently, that its functional derivative
$\, \mathslf{F}^{\,\prime}[\phi] \,$ at each~$\phi$ has compact
support, the sections $\delta\phi$, $\delta\phi_1^{}$, $\delta\phi_2^{}$
of~$V_\phi$ considered above may be allowed to have arbitrary support;
this is the case for local observables defined as integrals of local
densities over compact regions of space-time and, in particular, over
compact regions within a Cauchy surface $\Sigma$ (energy, momentum,
angular momentum, charges etc.\ within a finite volume).
Option~3: when $\mathslf{F}\,$ is \emph{local in time}, which we understand
to mean that its functional dependence on the fields is non-trivial only
within a time slice, or equivalently, that its functional derivative
$\, \mathslf{F}^{\,\prime}[\phi] \,$ at each~$\phi$ has temporally
compact support, we have to restrict the sections $\delta\phi$,
$\delta\phi_1^{}$, $\delta\phi_2^{}$ of $V_\phi$ considered above
to have spatially compact support (which can be achieved if the
sections $\phi_\lambda$, $\phi_{\lambda_1,\lambda_2}$ are supposed
to be independent of the parameters outside a spatially compact subset);
this is the case for global observables defined as integrals of local
densities over time slices and, in particular, over a Cauchy surface
$\Sigma$ (total energy, total momentum, total angular momentum, total
charges etc.).

Finally, covariant phase space $\mathscr{S}$ is defined to be the subset
of $\mathscr{C}$ consisting of the critical points of the action:
\begin{equation}
 \mathscr{S}~=~\{ \phi \smin \mathscr{C} \, / \, S^{\,\prime}[\phi] = 0 \}~.
\end{equation}
Formally, we can think of $\mathscr{S}$ as being a submanifold of
$\mathscr{C}$ whose tangent space at any point $\phi$ of $\mathscr{S}$
will be the subspace $T_\phi \mathscr{S}$ of the tangent space $T_\phi
\mathscr{C}$ consisting of the solu\-tions of the linearized equations
of motion (where ``linearized'' means ``linearized around the solution
$\phi$ of the full equations of motion''), which are precisely the
sections of $V_\phi$ belonging to the kernel of the corresponding
Jacobi operator $\; \mathscr{J}[\phi] : \Gamma(V_\phi) \longrightarrow
\Gamma(V_\phi^{\oast}) \;$:
\begin{equation}
 T_\phi \mathscr{S}~=~\ker \, \mathscr{J}[\phi]~.
\end{equation}

\subsection{Symplectic Structure}

Our next goal is to justify the term ``covariant phase space'' attributed to
$\mathscr{S}$ by showing that, formally, $\mathscr{S}$ carries a naturally
defined symplectic form $\Omega$, derived from an equally naturally defined
canonical form $\Theta$ by formal exterior differentiation. According to
Crnkovi\'c, Witten and Zuckerman \cite{CW,Cr,Zu} (see also \cite{Wo}), the
symplectic form $\Omega$ can be obtained by integration of a ``symplectic
current'', which is a closed $(n-1)$-form on space-time, over an arbitrary
spacelike hypersurface $\Sigma$. Here, we show that this ``symplectic
current'' can be derived directly from the multisymplectic form~$\omega$ or,
more explicitly, from the Poincar\'e\,-\,Cartan form $\omega_{\mathscr{L}}$ in
the Lagrangian approach and the De~Donder\,-\,Weyl form $\omega_{\mathscr{H}}$
in the Hamiltonian approach.

We begin with the definition of $\Theta$ and $\Omega$ in terms of $\theta$
and $\omega$, which is achieved by a mixture of contraction and pull-back:
given a point $\phi$ in $\mathscr{C}$ (a smooth section $\phi$ of $F$)
and smooth sections $\delta\phi$, $\delta\phi_1^{}$, $\delta\phi_2^{}$
of~$V_\phi$, insert $\delta\phi$ into the first of the $n$ arguments
of~$\theta$ or $\delta\phi_1^{}$ and $\delta\phi_2^{}$ into the first
two of the $n+1$ arguments of~$\omega$ and apply the definition of the
pull-back with~$\phi$ (which amounts to composition with the derivatives
$\partial\phi$ of~$\phi$) to the remaining $n-1$ arguments to obtain
$(n-1)$-forms on space-time which are integrated over $\Sigma$. Note
that these integrals exist if we assume that $\delta\phi$ and either
$\delta\phi_1^{}$ or $\delta\phi_2^{}$ have spatially compact support,
since this will intersect $\Sigma$ in a compact subset.

Explicitly, in the Lagrangian framework, we have
\begin{equation} \label{eq:THLAG1}
 \Theta_\phi(\delta\phi)~
 =~\int_\Sigma (\varphi,\partial\varphi)^\ast \,
   \theta_{\mathscr{L}}^{}(\delta\varphi,\partial\,\delta\varphi)
\end{equation}
and
\begin{equation} \label{eq:OMLAG1}
 \Omega_\phi(\delta\phi_1,\delta\phi_2)~
 =~\int_\Sigma (\varphi,\partial\varphi)^\ast \,
   \omega_{\mathscr{L}}^{}(\delta\varphi_1,\partial\,\delta\varphi_1\,,\,
                           \delta\varphi_2,\partial\,\delta\varphi_2)
\end{equation}
where the notation is the same as that employed in eq.~(\ref{eq:ACTN2}):
$\phi = \varphi \,$ is a section of $E$ over~$M$ and $j^1 \varphi =
(\varphi,\partial\varphi) \,$ is its (first) prolongation or derivative,
a section of $J^1 E$ over~$M$, while $\delta\phi = \delta\varphi$,
$\delta\phi_1^{} = \delta\varphi_1^{}$, $\delta\phi_2^{} = \delta
\varphi_2^{} \,$ are variations of $\, \phi = \varphi$, all sections
of $VE$ over $M$, and $\, \delta j^1 \varphi = (\delta\varphi,\partial\,
\delta\varphi)$, $\delta j^1 \varphi_1^{} = (\delta\varphi_1^{},\partial\,
\delta\varphi_1^{})$, $\delta j^1 \varphi_2^{} = (\delta\varphi_2^{},
\partial\,\delta\varphi_2^{}) \,$ are the induced variations of
$j^1 \varphi = (\varphi,\partial\varphi)$, all sections of
$\, V(J^1 E) \cong J^1(VE) \,$ over $M$. \linebreak
In local coordinates,
\[
 \delta\varphi~=~\frac{\partial}{\partial\lambda} \,
                 \varphi_\lambda \Big|_{\lambda=0}~
 =~\delta\varphi^i \, \frac{\partial}{\partial q^i}
\]
and
\[
 \delta j^1 \varphi~=~\frac{\partial}{\partial\lambda} \,
                      j^1 \varphi_\lambda \Big|_{\lambda=0}~
 =~\delta\varphi^i \, \frac{\partial}{\partial q^i} \, + \,
   \partial_\mu \delta\varphi^i \, \frac{\partial}{\partial q^i_\mu}
\]
whereas $\theta_{\mathscr{L}}^{}$ is given by eq.~(\ref{eq:MCF2}) and
$\omega_{\mathscr{L}}^{}$ by
\begin{eqnarray*}
 \omega_{\mathscr{L}}^{}
 &=& \Bigl( \, \frac{\partial^2 L}{\partial q^j \, \partial q^i_\mu} \;
               dq^i \,\smwedge\; dq^j \, + \,
               \frac{\partial^2 L}{\partial q^j_\nu \, \partial q^i_\mu} \;
               dq^i \,\smwedge\; dq^j_\nu \Bigr) \,\smwedge\;
     d^{\,n} x_\mu \\
 & & \mbox{} + \, \frac{\partial^2 L}{\partial x^\mu \, \partial q^i_\mu} \;
                  dq^i \,\smwedge\; d^{\,n} x \, - \,
     d \Bigl( L - \frac{\partial L}{\partial q^i_\mu} \, q^i_\mu \Bigr)
     \,\smwedge\; d^{\,n} x~.
\end{eqnarray*}
(The exterior derivative in the last term could be worked out explicitly,
but we shall not need this expression because the last two terms vanish
under contraction with two vertical vectors.) Then
\begin{equation} \label{eq:THLAG2}
 \Theta_\phi(\delta\phi)~
 =~\int_\Sigma d\sigma_\mu~
   \frac{\partial L}{\partial q^i_\mu}(\varphi,\partial\varphi) \;
   \delta\varphi^i
\end{equation}
and
\begin{equation} \label{eq:OMLAG2}
 \Omega_\phi(\delta\phi_1^{},\delta\phi_2^{})~
 =~\int_\Sigma d\sigma_\mu~J_\phi^\mu(\delta\phi_1^{},\delta\phi_2^{})
\end{equation}
with the ``symplectic current'' $J$ given by
\begin{equation} \label{eq:SCLAG1}
 \begin{array}{rcl}
  J_\phi^\mu(\delta\phi_1^{},\delta\phi_2^{})\!\!
  &=&\!\! {\displaystyle
           \frac{\partial^2 L}{\partial q^j \, \partial q_\mu^i}
           (\varphi,\partial\varphi) \;
           (\delta\varphi_1^i \; \delta\varphi_2^j -
            \delta\varphi_2^i \; \delta\varphi_1^j)} \\[4mm]
  & &     {\displaystyle + \;
           \frac{\partial^2 L}{\partial q_\nu^j \, \partial q_\mu^i}
           (\varphi,\partial\varphi) \;
           (\delta\varphi_1^i \; \partial_\nu^{} \delta\varphi_2^j -
            \delta\varphi_2^i \; \partial_\nu^{} \delta\varphi_1^j)}~,
 \end{array}
\end{equation}
or equivalently
\begin{equation} \label{eq:SCLAG2}
 \begin{array}{rcl}
  J_\phi^\mu(\delta\phi_1^{},\delta\phi_2^{})\!\!
  &=&\!\! {\displaystyle \mbox{} - \,
           \Big( \, \frac{\partial^2 L}{\partial q^j \, \partial q_\mu^i}
                    (\varphi,\partial\varphi) \;
                    \delta\varphi_1^j \, + \,
                    \frac{\partial^2 L}{\partial q_\nu^j \, \partial q_\mu^i}
                    (\varphi,\partial\varphi) \;
                    \partial_\nu^{} \delta\varphi_1^j \, \Big) \;
           \delta\varphi_2^i} \\[4mm]
  & &\!\! {\displaystyle \mbox{} + \,
           \Big( \, \frac{\partial^2 L}{\partial q^j \, \partial q_\mu^i}
                    (\varphi,\partial\varphi) \;
                    \delta\varphi_2^j \, + \,
                    \frac{\partial^2 L}{\partial q_\nu^j \, \partial q_\mu^i}
                    (\varphi,\partial\varphi) \;
                    \partial_\nu^{} \delta\varphi_2^j \, \Big) \;
           \delta\varphi_1^i}~.
 \end{array}
\end{equation}

The same results can be obtained even more directly in the Hamiltonian
framework, in which we have
\begin{equation} \label{eq:THHAM1}
 \Theta_\phi(\delta\phi)~
 =~\int_\Sigma (\varphi,\pi)^\ast \,
   \theta_\mathscr{H}(\delta\varphi,\delta\pi)
\end{equation}
and
\begin{equation} \label{eq:OMHAM1}
 \Omega_\phi(\delta\phi_1^{},\delta\phi_2^{})~
 =~\int_\Sigma (\varphi,\pi)^\ast \,
   \omega_\mathscr{H} (\delta\varphi_1^{},\delta\pi_1^{}\,,\,
                       \delta\varphi_2^{},\delta\pi_2^{})
\end{equation}
where the notation is the same as that employed in eq.~(\ref{eq:ACTN3}):
$\phi = (\varphi,\pi) \,$ is a section of $\vec{J}^{\,1\,\oast} E$
over~$M$ while $\delta\phi = (\delta\varphi,\delta\pi)$, $\delta\phi_1^{}
= (\delta\varphi_1^{},\delta\pi_1^{})$, $\delta\phi_2^{} = (\delta\varphi_2^{},
\delta\pi_2^{}) \,$ are variations of $\, \phi = (\varphi,\pi)$, all
sections of $V(\vec{J}^{\,1\,\oast} E)$ over $M$. In local coordinates,
$$
 \delta\varphi~
 =~\frac{\partial}{\partial \lambda} \, \varphi_\lambda \Big|_{\lambda=0}~
 =~\delta\varphi^i \, \frac{\partial}{\partial q^i}~~~,~~~
 \delta\pi~
 =~\frac{\partial}{\partial \lambda} \, \pi_\lambda \Big|_{\lambda=0}~
 =~\delta\pi^\mu_i \, \frac{\partial}{\partial p^\mu_i}
$$
whereas $\theta_{\mathscr{H}}^{}$ is given by eq.~(\ref{eq:MCF3}) and
$\omega_{\mathscr{H}}^{}$ by
$$
 \omega_\mathscr{H}~
 =~dq^i \,\smwedge\; dp\>\!_i^\mu \,\smwedge\; d^{\,n} x_\mu \, - \,
   dH \,\smwedge\; d^{\,n} x
$$
(The exterior derivative in the last term could be worked out explicitly,
but we shall not need this expression because the last term vanishes
under contraction with two vertical vectors.) Then
\begin{equation} \label{eq:THHAM2}
 \Theta_\phi(\delta\phi)~
 = \, \int_\Sigma d\sigma_\mu \; \pi^\mu_i \; \delta\varphi^i
\end{equation}
and
\begin{equation} \label{eq:OMHAM2}
 \Omega_\phi(\delta\phi_1^{},\delta\phi_2^{})~
 =~\int_\Sigma d\sigma_\mu~J_\phi^\mu(\delta\phi_1^{},\delta\phi_2^{})
\end{equation}
with the ``symplectic current'' $J$ given by
\begin{equation} \label{eq:SCHAM}
 J_\phi^\mu(\delta\phi_1^{},\delta\phi_2^{})~
 =~\delta\varphi_1^i \; \delta\pi_{2,i}^{\hphantom{2,}\mu} \, - \,
   \delta\varphi_2^i \; \delta\pi_{1,i}^{\hphantom{1,}\mu}~.
\end{equation}
Incidentally, these formulas show that, just like in mechanics,
the canonical form $\Theta$ and the symplectic form $\Omega$
do not depend on the choice of the Hamiltonian $\mathscr{H}$.

Another important result, duly emphasized in the literature~%
\cite{CW,Cr,Zu,Wo}, is the fact that on covariant phase space
$\mathscr{S}$, the symplectic form  $\Omega$ does not depend on
the choice of the hypersurface $\Sigma$ used in its definition,
since for any solution $\phi$ of the equations of motion and any
two solutions $\delta\phi_1^{}$, $\delta\phi_2^{}$ of the linearized
equations of motion, the ``symplectic current'' $J_\phi(\delta\phi_1^{},
\delta\phi_2^{})$ is a closed form on space-time. To prove this, assume
that $\phi$ is a point in $\mathscr{S}$ and observe that a tangent
vector $\delta\phi$ in $T_\phi \mathscr{C}$ belongs to the subspace
$T_\phi \mathscr{S}$ if and only if $\delta\phi$, as a section of
$V_\phi$, satisfies the pertinent Jacobi equation, which reads
\begin{equation}
 \begin{array}{l}
  {\displaystyle
   \partial_\mu \Big( \, \frac{\partial^2 L}{\partial q^j \, \partial q^i_\mu}
                         (\varphi,\partial\varphi) \; \delta\varphi^j \, + \,
                         \frac{\partial^2 L}
                              {\partial q^j_\nu \, \partial q^i_\mu}
                         (\varphi,\partial\varphi) \;
                         \partial_\nu^{} \delta\varphi^j \, \Big)} \\[5mm]
  \quad =~{\displaystyle
           \frac{\partial^2 L}{\partial q^j \, \partial q^i}
           (\varphi,\partial\varphi) \; \delta\varphi^j \, + \,
           \frac{\partial^2 L}{\partial q^i \, \partial q_\nu^j}
           (\varphi,\partial\varphi) \; \partial_\nu^{} \delta\varphi^j}
 \end{array}
\end{equation}
in the Lagrangian framework and
\begin{equation}
 \begin{array}{c}
  \partial_\mu \, \delta\pi^\mu_i~
  =~{\displaystyle
     \frac{\partial^{\>\!2} H}{\partial q^j \, \partial q^i}
     (\varphi,\pi) \; \delta\varphi^j \, + \,
     \frac{\partial^{\>\!2} H}{\partial p\>\!_j^\nu \, \partial q^i}
     (\varphi,\pi) \; \delta \pi_j^\nu} \\[5mm]
  \partial_\mu \, \delta \varphi^i~
  = \; {\displaystyle -\,
     \frac{\partial^{\>\!2} H}{\partial q^j \, \partial p\>\!_i^\mu}
     (\varphi,\pi) \; \delta\varphi^j \, - \,
     \frac{\partial^{\>\!2} H}{\partial p\>\!_j^\nu \, \partial p\>\!_i^\mu}
     (\varphi,\pi) \; \delta \pi_j^\nu}
 \end{array}
\end{equation}
in the Hamiltonian framework. Thus if $\delta\phi_1^{}$ and $\delta\phi_2^{}$
both satisfy the Jacobi equation, we have
\begin{eqnarray*}
 \partial_\mu J_\phi^\mu(\delta\phi_1^{},\delta\phi_2^{}) \!\!
 &=&\!\! \mbox{} - \, \partial_\mu
         \Big( \, \frac{\partial^2 L}{\partial q^j \, \partial q^i_\mu}
                  (\varphi,\partial\varphi) \; \delta\varphi_1^j \, + \,
                  \frac{\partial^2 L}{\partial q^j_\nu \, \partial q^i_\mu}
                  (\varphi,\partial\varphi) \;
                  \partial_\nu \delta\varphi_1^j \, \Big) \;
         \delta\varphi_2^i                                              \\
 & &\!\! \mbox{} - \,
         \Big( \, \frac{\partial^2 L}{\partial q^j \, \partial q^i_\mu}
                  (\varphi,\partial\varphi) \; \delta\varphi_1^j \, + \,
                  \frac{\partial^2 L}{\partial q^j_\nu \, \partial q^i_\mu}
                  (\varphi,\partial\varphi) \;
                  \partial_\nu\delta\varphi_1^j \, \Big) \;
         \partial_\mu \delta\varphi_2^i                                 \\
 & &\!\! \mbox{} + \, \partial_\mu
         \Big( \, \frac{\partial^2 L}{\partial q^j \, \partial q^i_\mu}
                  (\varphi,\partial\varphi) \; \delta\varphi_2^j \, + \,
                  \frac{\partial^2 L}{\partial q^j_\nu \, \partial q^i_\mu}
                  (\varphi,\partial\varphi) \;
                  \partial_\nu \delta\varphi_2^j \, \Big) \;
         \delta\varphi_1^i                                              \\
 & &\!\! \mbox{} + \,
         \Big( \, \frac{\partial^2 L}{\partial q^j \, \partial q^i_\mu}
                  (\varphi,\partial\varphi) \; \delta\varphi_2^j \, + \,
                  \frac{\partial^2 L}{\partial q^j_\nu \, \partial q^i_\mu}
                  (\varphi,\partial\varphi) \;
                  \partial_\nu\delta\varphi_2^j \, \Big) \;
                  \partial_\mu \delta\varphi_1^i
\end{eqnarray*}
\begin{eqnarray*}
 \phantom{\partial_\mu J_\phi^\mu(\delta\phi_1^{},\delta\phi_2^{}) \!\!}
 &=&\!\! \mbox{} - \,
         \Big( \, \frac{\partial^2 L}{\partial q^j \, \partial q^i}
                  (\varphi,\partial\varphi) \; \delta\varphi_1^j \, + \,
                  \frac{\partial^2 L}{\partial q^i \partial q_\nu^j}
                  (\varphi,\partial\varphi) \;
                  \partial_\nu \delta\varphi_1^j \, \Big) \;
         \delta\varphi_2^i                                              \\
 & &\!\! \mbox{} - \,
         \Big( \, \frac{\partial^2 L}{\partial q^j \, \partial q^i_\mu}
                  (\varphi,\partial\varphi) \; \delta\varphi_1^j \, + \,
                  \frac{\partial^2 L}{\partial q^j_\nu \, \partial q^i_\mu}
                  (\varphi,\partial\varphi) \;
                  \partial_\nu\delta\varphi_1^j \, \Big) \;
         \partial_\mu \delta\varphi_2^i                                 \\
 & &\!\! \mbox{} + \,
         \Big( \, \frac{\partial^2 L}{\partial q^j \, \partial q^i}
                  (\varphi,\partial\varphi) \; \delta\varphi_2^j \, + \,
                  \frac{\partial^2 L}{\partial q^i \partial q_\nu^j}
                  (\varphi,\partial\varphi) \;
                  \partial_\nu \delta\varphi_2^j \, \Big) \;
         \delta\varphi_1^i                                              \\
 & &\!\! \mbox{} + \,
         \Big( \, \frac{\partial^2 L}{\partial q^j \, \partial q^i_\mu}
                  (\varphi,\partial\varphi) \; \delta\varphi_2^j \, + \,
                  \frac{\partial^2 L}{\partial q^j_\nu \, \partial q^i_\mu}
                  (\varphi,\partial\varphi) \;
                  \partial_\nu\delta\varphi_2^j \, \Big) \;
         \partial_\mu \delta\varphi_1^i
\end{eqnarray*}
in the Lagrangian framework and
\begin{eqnarray*}
 \partial_\mu J_\phi^\mu(\delta\phi_1^{},\delta\phi_2^{}) \!\!
 &=&\!\! \partial_\mu \delta\varphi_1^i \;
         \delta\pi_{2,i}^{\hphantom{2,}\mu} \, + \,
         \delta\varphi_1^i \;
         \partial_\mu \delta\pi_{2,i}^{\hphantom{2,}\mu} \, - \,
         \partial_\mu \delta\varphi_2^i \;
         \delta\pi_{1,i}^{\hphantom{1,}\mu} \, - \,
         \delta\varphi_2^i \;
         \partial_\mu \delta\pi_{1,i}^{\hphantom{1,}\mu}                \\[2mm]
 &=& -\, \Big( \, \frac{\partial^{\>\!2} H}
                       {\partial q^j \, \partial p\>\!_i^\mu}
                  (\varphi,\pi) \; \delta\varphi_1^j \, + \,
                  \frac{\partial^{\>\!2} H}
                       {\partial p\>\!_j^\nu \, \partial p\>\!_i^\mu}
                  (\varphi,\pi) \;
                  \delta\pi_{1,j}^{\hphantom{1,}\nu} \, \Big) \;
         \delta\pi_{2,i}^{\hphantom{2,}\mu}                             \\
 & & +~  \delta\varphi_1^i \;
         \Big( \, \frac{\partial^{\>\!2} H}{\partial q^j \, \partial q^i}
                  (\varphi,\pi) \; \delta\varphi_2^j \, + \,
                  \frac{\partial^{\>\!2} H}
                       {\partial p\>\!_j^\nu \, \partial q^i}
                  (\varphi,\pi) \;
                  \delta\pi_{2,j}^{\hphantom{2,}\nu} \, \Big)           \\
 & & +\, \Big( \, \frac{\partial^{\>\!2} H}
                       {\partial q^j \, \partial p\>\!_i^\mu}
                  (\varphi,\pi) \; \delta\varphi_2^j \, + \,
                  \frac{\partial^{\>\!2} H}
                       {\partial p\>\!_j^\nu \, \partial p\>\!_i^\mu}
                  (\varphi,\pi) \;
                  \delta\pi_{2,j}^{\hphantom{2,}\nu} \, \Big) \;
         \delta\pi_{1,i}^{\hphantom{1,}\mu}                             \\
 & & -~  \delta\varphi_2^i \;
         \Big( \, \frac{\partial^{\>\!2} H}{\partial q^j \, \partial q^i}
                  (\varphi,\pi) \; \delta\varphi_1^j \, + \,
                  \frac{\partial^{\>\!2} H}
                       {\partial p\>\!_j^\nu \, \partial q^i}
                  (\varphi,\pi) \;
                  \delta\pi_{1,j}^{\hphantom{1,}\nu} \, \Big)
\end{eqnarray*}
in the Hamiltonian framework: obviously, both of these expressions vanish.

Of course, independence of the choice of hypersurface holds only for $\Omega$
but not for~$\Theta$. In fact, if $M_{1,2}$ is a region of space-time whose
boundary is the disjoint union of two hypersurfaces $\Sigma_1$ and $\Sigma_2$,
then $\, \Omega_{\Sigma_2} = \Omega_{\Sigma_1} \,$ but
\begin{equation}
 \Theta_{\Sigma_2} - \Theta_{\Sigma_1}~=~\delta S_{M_{1,2}}
\end{equation}
where $S_{M_{1,2}}$ is the action calculated by integration over $M_{1,2}$
and $\delta$ is the functional exterior derivative, or variational derivative,
on $\mathscr{S}$.

\subsection{Poisson Bracket} 

Given a relativistic field theory with a regular first-order Lagrangian,
one expects each of the corresponding Jacobi operators $\mathscr{J}[\phi]$
($\phi \smin \mathscr{S}$) to form a hyperbolic system of second-order
partial differential operators. A typical example is provided by the sigma
model, where $E$ is a trivial product bundle $M \times Q$, with a given
Lorentzian metric $g$ on the base manifold $M$, as usual, and a given
Riemannian metric $h$ on the typical fiber $Q$. Its Lagrangian is
$$
 L~=~{\textstyle \frac{1}{2}} \, \sqrt{|g|} \,
     g^{\mu\nu} \, h_{ij} \, q^i_\mu \, q^j_\nu~,
$$
so that the coefficients of the highest degree terms of the
Jacobi operator $\mathscr{J}[\varphi]$ are
$$
 \frac{\partial^2 L}{\partial q_\nu^j \, \partial q_\mu^i} \,
 (\varphi,\partial\varphi)~
 =~{\textstyle \frac{1}{2}} \, \sqrt{|g|} \, g^{\mu\nu} \, h_{ij}(\varphi)~,
$$
which clearly exhibits the hyperbolic nature of the resulting linearized
field equations.

A general feature of hyperbolic systems of linear partial differential
equations is the possibility to guarantee existence and uniqueness of
various types of Green functions. In the present context, what we need
is existence and uniqueness of the retarded Green function $G_\phi^-$,
the advanced Green function $G_\phi^+$ and the causal Green function
$G_\phi$ for the Jacobi operator $\mathscr{J}[\phi]$, for each
$\, \phi \smin \mathscr{S}$. By definition, the first two are
solutions of the inhomogeneous Jacobi equations
\begin{equation}
 \mathscr{J}_x[\phi] \; G_\phi^{\pm}(x,y)~=~\delta(x,y) \quad , \quad
 \mathscr{J}_y[\phi] \; G_\phi^{\pm}(x,y)~=~\delta(x,y)~,
\end{equation}
or more explicitly,
\begin{equation}
 \mathscr{J}_x[\phi]_{km}^{} \, G_\phi^{\pm \, ml}(x,y)~
 =~\delta_k^l \, \delta(x,y) \quad , \quad
 \mathscr{J}_y[\phi]_{km}^{} \, G_\phi^{\pm \, lm}(x,y)~
 =~\delta_k^l \, \delta(x,y)~,
\end{equation}
where $\mathscr{J}_z[\phi]$ denotes the Jacobi operator with respect to the
variable $z$, characterized by the following support condition: for any
$\, x,y \smin M$, $G^-_\phi(x,y) = 0 \,$ when $\, x \nsmin J^+(y) \,$
and $G_\phi^+(x,y) = 0 \,$ when $x \nsmin J^-(y)$, where $J^+(y)$ and
$J^-(y)$ are the future cone and the past cone of $y$, respectively.
The causal Green function, also called the propagator, is then simply
their difference:
\begin{equation} \label{eq:CRAGF}
 G_\phi~=~G_\phi^- - \, G_\phi^+~.
\end{equation}
Obviously, it satisfies the homogeneous Jacobi equations
\begin{equation}
 \mathscr{J}_x[\phi] \; G_\phi(x,y)~=~0 \quad , \quad
 \mathscr{J}_y[\phi] \; G_\phi(x,y)~=~0~.
\end{equation}
Note that the symmetry of the Jacobi operator $\mathscr{J}[\phi]$, stemming
from the fact that it represents the second variational derivative of the
action, forces these Green functions to satisfy the following exchange and
symmetry properties:
\begin{equation}
 G_\phi^{\pm \, lk}(y,x)~=~G_\phi^{\mp \, kl}(x,y) \quad , \quad
 G_\phi^{lk}(y,x)~= \; - \, G_\phi^{kl}(x,y)~.
\end{equation}
It should be pointed out that existence and uniqueness of these Green
functions cannot be guaranteed in complete generality: this requires
not only that~$M$ be globally hyperbolic but also that the linearized
field equations should form a hyperbolic system. Here, we shall simply
assume this to be the case and proceed from there; further comments on
the question will be deferred to the end of the section.

Our next step will be to study certain specific (distributional) solutions
$\mathslf{X}_{\mathsmf{F}}^{}[\phi]$ of the general inhomogeneous Jacobi
equation
\begin{equation}
 \mathscr{J}[\phi](\mathslf{X}_{\mathsmf{F}}^{}[\phi])~
 =~\mathslf{F}^{\,\prime}[\phi]
\end{equation}
for smooth functionals $\mathslf{F}$ on covariant phase space which are
(at least) local in time. To~eliminate the ambiguity in this equation
stemming from the fact that the functional derivative $\mathslf{F}^%
{\,\prime}[\phi]$ on its rhs belongs to the space $T_\phi^\ast \mathscr{S}$
which is only a quotient space of the image space $T_\phi^\ast \mathscr{C}$
of the Jacobi operator $\mathscr{J}[\phi]$ (an inclusion of the form
$\, T_\phi \mathscr{S} \,\smsubset T_\phi \mathscr{C}$ \linebreak
induces a natural projection from $T_\phi^\ast \mathscr{C}$ to
$T_\phi^\ast \mathscr{S}$), it is necessary to first of all extend
the given functional $\mathslf{F}$ on $\mathscr{S}$ to a functional
$\,\tilde{\!\mathslf{F}\,}$ on $\mathscr{C}$ of the same type
(smooth and local in time), whose functional derivative
$\, \tilde{\!\!\mathslf{F}}\vphantom{f}^{\,\prime}[\phi]$ 
does belong to the space $T_\phi^\ast \mathscr{C}$ which, as we
recall from eq.~(\ref{eq:TCS3}), consists of the distributional
sections of $V_\phi^{\oast}$ of temporally compact support.
Next, convolution with the retarded and advanced Green function
introduced above produces formal vector fields over~$\mathscr{S}$
which to each solution $\, \phi \smin \mathscr{S} \,$ of
the field equations associate (distributional) sections
$\, \mathslf{X}_{\,\,\,\tilde{\!\!\mathsmf{F}}}^{\,-}[\phi]$ and
$\, \mathslf{X}_{\,\,\,\tilde{\!\!\mathsmf{F}}}^{\,+}[\phi]$ of~%
$V_\phi$, respectively. In~local coordinates, their definition
can (formally and at least when the intersection of the two
supports is contained in the coordinate domain) be written
in the form
\begin{equation} \label{eq:CONV1}
 \mathslf{X}_{\,\,\,\tilde{\!\!\mathsmf{F}}}^{\,\pm}[\phi]^{\,k}(x)~
 =~\int_M d^{\,n} y~G_\phi^{\pm \, kl}(x,y) \,
   \frac{\delta \,\,\tilde{\!\!\mathslf{F}}}{\delta\phi^{\,l}}[\phi](y)~.
\end{equation}
Both of them satisfy the inhomogeneous Jacobi equation
\begin{equation} \label{eq:JEQINH}
 \mathscr{J}[\phi](\mathslf{X}_{\,\,\,\tilde{\!\!\mathsmf{F}}}^{\,\pm}[\phi])~
 =~\,\,\tilde{\!\!\mathslf{F}}\vphantom{f}^{\,\prime}[\phi]~.
\end{equation}
Similarly, convolution with the causal Green function leads to another
formal vector field over~$\mathscr{S}$ which to each solution $\, \phi
\smin \mathscr{S} \,$ of the field equations associates a (distributional)
section $\mathslf{X}_{\,\,\tilde{\!\!\mathsmf{F}}}^{}[\phi]$ of~$V_\phi$.
Again, in local coordinates, its definition can (formally and at least
when the intersection of the two supports is contained in the coordinate
domain) be written in the form
\begin{equation} \label{eq:CONV2}
 \mathslf{X}_{\,\,\tilde{\!\!\mathsmf{F}}}^{}[\phi]^{\,k}(x)~
 =~\int_M d^{\,n} y~G_\phi^{kl}(x,y) \,
   \frac{\delta \,\,\tilde{\!\!\mathslf{F}}}{\delta\phi^{\,l}}[\phi](y)~.
\end{equation}
It satisfies the homogeneous Jacobi equation
\begin{equation} \label{eq:JEQHOM}
 \mathscr{J}[\phi](\mathslf{X}_{\,\,\tilde{\!\!\mathsmf{F}}}^{}[\phi])~=~0~,
\end{equation}
since according to eq.~(\ref{eq:CRAGF})
\begin{equation} \label{eq:CRASL}
 \mathslf{X}_{\,\,\tilde{\!\!\mathsmf{F}}}^{}[\phi]~
 =~\mathslf{X}_{\,\,\,\tilde{\!\!\mathsmf{F}}}^{\,-}[\phi] -
   \mathslf{X}_{\,\,\,\tilde{\!\!\mathsmf{F}}}^{\,+}[\phi]~.
\end{equation}
Note that the convolutions in eqs~(\ref{eq:CONV1}) and~(\ref{eq:CONV2}) exist
due to our support assumptions on $\,\, \tilde{\!\!\mathslf{F}}$ (requiring
$\,\,\, \tilde{\!\!\mathslf{F}}\vphantom{f}^{\,\prime}[\phi] \,$ to have
temporally compact support) and due to the support properties of the Green
functions $G_\phi^\pm$ and $G_\phi^{}$.

According to eq.~(\ref{eq:JEQHOM}), the prescription of associating to
each solution $\, \phi \smin \mathscr{S} \,$ of the field equations the
section $\, \mathslf{X}_{\,\,\tilde{\!\!\mathsmf{F}}}^{}[\phi] \,$
of~$V_\phi$ defines a formal vector field on~$\mathscr{S} \,$ which is
tangent to $\mathscr{S} \,$. (It becomes more than just a formal vector
field if $\,\,\tilde{\!\!\mathslf{F}}$ is such that $\, \mathslf{X}_%
{\,\,\tilde{\!\!\mathsmf{F}}}^{}[\phi] \,$ belongs to $T_\phi \mathscr{S} \,$,
which requires it to be not just a distributional section but a smooth section
of~$V_\phi$ and to satisfy appropriate support properties; we shall come back
to this point later on.) \linebreak The main statement about this formal vector
field, to be proved below, is that (a) it does not depend on the choice of the
extension $\,\,\tilde{\!\!\mathslf{F}}$ of $\mathslf{F}$, so we may simply
denote it by $\, \mathslf{X}_{\mathsmf{F}}^{}[\phi]$, and (b) that it is
formally the Hamiltonian vector field associated to~$\mathslf{F}$ with
respect to the symplectic form $\Omega$ discussed in the previous subsection.
More explicitly, we claim that for any solution $\, \phi \smin \mathscr{S} \,$
of the field equations and any smooth section $\delta\phi$ of~$V_\phi$ with
spatially compact support which is a solution of the linearized field
equations, we have
\phantom{this is just to guarantee correct vertical spacing}
\begin{equation} \label{eq:HAMVF2}
 \Omega_\phi(\mathslf{X}_{\mathsmf{F}}^{}[\phi],\delta\phi)~
 =~\mathslf{F}^{\,\prime}[\phi] \cdot \delta\phi~.
\end{equation}
Note that under the assumptions stated, both sides of this equation make
sense although we have originally defined $\Omega_\phi(\delta\phi_1^{},
\delta\phi_2^{})$ only in the case where both $\delta\phi_1^{}$ and
$\delta\phi_2^{}$ are smooth; the extension of this definition, given
in the previous subsection, to the case where one of them is a distribution
is straightforward.

To prove this key statement, let us begin by recalling that the symplectic
form $\Omega$ and the symplectic current $J$ of the previous subsection are
really defined on $\mathscr{C}$ and not only on $\mathscr{S}$~-- the only
difference is that on $\mathscr{C}$, $\Omega$ is only a presymplectic form
so that $J$ should be more appropriately called the presymplectic current
and that $J$ on $\mathscr{C}$ is no longer be conserved so that $\Omega$
on $\mathscr{C}$ will depend on the choice of the hypersurface~$\Sigma$.
At any rate, we can almost literally repeat the calculation performed
at the end of the previous subsection, either in the Lagrangian or in
the Hamiltonian formulation, to show that for any solution $\, \phi
\smin \mathscr{S} \,$ of the field equations and any smooth section
$\delta\phi$ of~$V_\phi$ with spatially compact support, we have
\begin{equation} \label{eq:DIVSC1}
 \partial_\mu \;\! J_\phi^\mu
 (\mathslf{X}_{\,\,\,\tilde{\!\!\mathsmf{F}}}^{\,\pm}[\phi],\delta\phi)~
 =~(\, \mathscr{J}[\phi]_{kl}^{} \,
       \mathslf{X}_{\,\,\,\tilde{\!\!\mathsmf{F}}}^{\,\pm}[\phi]^{\,l}) \;
   \delta\phi^{\,k} \, - \,
   (\, \mathscr{J}[\phi]_{kl}^{} \, \delta\phi^{\,l}) \;
   \mathslf{X}_{\,\,\,\tilde{\!\!\mathsmf{F}}}^{\,\pm}[\phi]^{\,k}~,
\end{equation}
so that if $\delta\phi$ is a solution of the linearized field equations,
\begin{equation} \label{eq:DIVSC2}
 \partial_\mu \;\! J_\phi^\mu
 (\mathslf{X}_{\,\,\,\tilde{\!\!\mathsmf{F}}}^{\,\pm}[\phi],\delta\phi)~
 =~\frac{\delta \,\,\tilde{\!\!\mathslf{F}}}{\delta\phi}[\phi] \cdot
   \delta\phi~.
\end{equation}
Now since, by assumption, the support of $\, (\delta\,\,\tilde{\!\!\mathslf{F}}
/ \delta\phi) [\phi] \,$ is contained in some time slice, we can choose two
Cauchy surfaces $\Sigma_-$ to the past and $\Sigma_+$ to the future of this
time slice and, using that $\delta\phi$ has spatially compact support,
integrate eq.~(\ref{eq:DIVSC2}) over the time slice~$S_-$ between
$\Sigma_-$ and~$\Sigma$ and similarly over the time slice~$S_+$
between $\Sigma$ and~$\Sigma_+$. Applying Stokes' theorem, this gives
\begin{eqnarray*}
 \int_\Sigma d\sigma_\mu(x) \,
 J_\phi^\mu(\mathslf{X}_{\,\,\,\tilde{\!\!\mathsmf{F}}}^{\,-}[\phi],
            \delta\phi)(x) \!\!
 &=&\!\!\!
 \int_{\Sigma_-} \!\! d\sigma_\mu(x) \,
 J_\phi^\mu(\mathslf{X}_{\,\,\,\tilde{\!\!\mathsmf{F}}}^{\,-}[\phi],
            \delta\phi)(x) \, +
 \int_{S_-} \!\! d^{\,n} x~
 \frac{\delta \,\,\tilde{\!\!\mathslf{F}}}{\delta\phi}[\phi](x) \cdot
 \delta\phi(x) \, , \\
 \int_\Sigma d\sigma_\mu(x) \,
 J_\phi^\mu(\mathslf{X}_{\,\,\,\tilde{\!\!\mathsmf{F}}}^{\,+}[\phi],
            \delta\phi)(x) \!\!
 &=&\!\!\!
 \int_{\Sigma_+} \!\! d\sigma_\mu(x) \,
 J_\phi^\mu(\mathslf{X}_{\,\,\,\tilde{\!\!\mathsmf{F}}}^{\,+}[\phi],
            \delta\phi)(x) \, +
 \int_{S_+} \!\! d^{\,n} x~
 \frac{\delta \,\,\tilde{\!\!\mathslf{F}}}{\delta\phi}[\phi](x) \cdot
 \delta\phi(x) \, .
\end{eqnarray*}
But the support conditions on $G_\phi^\pm$, together with the fact that
the support of $\, (\delta\,\,\tilde{\!\!\mathslf{F}} / \delta\phi)[\phi]$
\linebreak
lies to the future of~$\Sigma_-$ and to the past of~$\Sigma_+$, imply that
$\mathslf{X}_{\,\,\,\tilde{\!\!\mathsmf{F}}}^{\,-}[\phi]$ vanishes on~%
$\Sigma_-$ and similarly that $\mathslf{X}_{\,\,\,\tilde{\!\!\mathsmf{F}}}^%
{\,+}[\phi]$ vanishes on~$\Sigma_+$, so the first term on the rhs of each
of these equations is zero. Thus taking their difference and inserting
eq.~(\ref{eq:CRASL}), we get
$$
 \int_\Sigma d\sigma_\mu(x) \,
 J_\phi^\mu(\mathslf{X}_{\,\,\tilde{\!\!\mathsmf{F}}}^{}[\phi],
            \delta\phi)(x)~
 =~\int_{S_-} \!\! d^{\,n} x~
   \frac{\delta \,\,\tilde{\!\!\mathslf{F}}}{\delta\phi}[\phi](x) \cdot
   \delta\phi(x) \, + \,
   \int_{S_+} \!\! d^{\,n} x~
   \frac{\delta \,\,\tilde{\!\!\mathslf{F}}}{\delta\phi}[\phi](x) \cdot
   \delta\phi(x)~,
$$
and since $\, (\delta\,\,\tilde{\!\!\mathslf{F}} / \delta\phi)[\phi] \,$
vanishes outside $\, S_- \smcup S_+$,
\begin{equation} \label{eq:HAMVF3}
 \Omega_\phi(\mathslf{X}_{\,\,\tilde{\!\!\mathsmf{F}}}^{}[\phi],\delta\phi)~
 =~\int_M d^{\,n} x~
   \frac{\delta \,\,\tilde{\!\!\mathslf{F}}}{\delta\phi}[\phi](x) \cdot
   \delta\phi(x)~.
\end{equation}
Finally, observe that since $\delta\phi$ is supposed to be a solution
of the linearized field equations (and hence tangent to $\mathscr{S}$),
the rhs of this equation does not depend on the choice of the extension 
$\,\,\tilde{\!\!\mathslf{F}}$ of~$\mathslf{F}$. Therefore,
$\mathslf{X}_{\,\,\tilde{\!\!\mathsmf{F}}}^{}[\phi]$ will not
depend on this choice either provided the symplectic form
$\Omega_\phi$ is weakly non-degenerate. Now using the space-time
split of~$M$ over~$\Sigma$ provided by the tangent vector field
$\partial_t^{}$ of some global time function $t$ on $M$ \linebreak
or its dual $dt$, and identifying solutions $\delta\phi$ of the
linearized field equations with their Cauchy data on $\Sigma$,%
\footnote{Explicitly, in the Lagrangian formalism, the Cauchy data for
$\delta\varphi$ on~$M$ are $\delta\varphi$ and $\delta\dot{\varphi}$
on~$\Sigma$, whereas in the Hamiltonian formalism, the Cauchy data for
$(\delta\varphi,\delta\pi)$ on~$M$ are $\delta\varphi$ and $\delta\pi^0$
on~$\Sigma$.}
it can be seen by direct inspection, either of eqs~(\ref{eq:OMLAG2}) and~%
(\ref{eq:SCLAG1}) in the Lagrangian formalism or of eqs~(\ref{eq:OMHAM2})
and~(\ref{eq:SCHAM}) in the Hamiltonian formalism, that the expression
$\Omega_\phi(\delta\phi_1,\delta\phi_2)$ can only be zero for all
$\delta\phi_2$ if $\delta\phi_1$ vanishes, as soon as we require
the Lagrangian $L$ to be regular in time derivatives, that is,
to satisfy
\begin{equation}
 \det \frac{\partial^{\>\!2} L}
           {\partial q_0^{i\vphantom{j}} \, \partial q_0^j}~\neq~0~,
\end{equation}
or equivalently, the Hamiltonian to be regular in timelike conjugate
momenta, that is, to satisfy
\begin{equation}
 \det \frac{\partial^{\>\!2} H}
           {\partial p\>\!_i^0 \, \partial p\>\!_j^0}~\neq~0~.
\end{equation}
Moreover, it can be shown that this statement will remain true if $\delta
\phi_1$ is allowed to be a distributional solution of the linearized field
equations with arbitrary support, as long as $\delta\phi_2$ runs through
the space of smooth solutions of the linearized field equations with
spatially compact support.

Let us summarize this fundamental result in the form of a theorem.
\begin{The}
 With respect to the symplectic form $\Omega$ on covariant phase space as
 defined by Crnkovi\'c, Witten and Zuckerman, the Hamiltonian vector field
 $\mathslf{X}_{\mathsmf{F}}^{}$ associated with a functional $\mathslf{F}$
 which is local in time is given by convolution of the functional derivative
 of $\mathslf{F}$ with the causal Green function of the corresponding Jacobi
 operator.
\end{The}
Note that in view of the regularity conditions employed to arrive at this
conclusion, the previous construction does not apply directly to degenerate
systems such as gauge theories: these require a separate treatment.

Having established eq.~(\ref{eq:HAMVF2}), it is now easy to write down
the Poisson bracket of two functionals $\mathslf{F}$ and $\mathslf{G}\,$
on $\mathscr{S} \,$: it is, in complete analogy with eq.~(\ref{eq:POISB1}),
given by
\begin{equation} \label{eq:POISB2}
 \{ \mathslf{F} , \mathslf{G} \} [\phi]~
 =~\mathslf{F}^{\,\prime}[\phi] \cdot \mathslf{X}_{\mathsmf{G}}^{}[\phi]~
 = \; - \,
   \mathslf{G}^{\,\prime}[\phi] \cdot \mathslf{X}_{\mathsmf{F}}^{}[\phi]~,
\end{equation}
or
\begin{equation} \label{eq:POISB3}
 \{ \mathslf{F} , \mathslf{G} \} [\phi]~
 =~\int_M d^{\,n} x~\frac{\delta \mathslf{F}}{\delta\phi^{\,k}}[\phi](x) \;
   \mathslf{X}_{\mathsmf{G}}^{}[\phi]^{\,k}(x)~
 = \; - \,
   \int_M d^{\,n} x~\frac{\delta \mathslf{G}}{\delta\phi^{\,k}}[\phi](x) \;
   \mathslf{X}_{\mathsmf{F}}^{}[\phi]^{\,k}(x)~.
\end{equation}
Inserting eq.~(\ref{eq:CONV2}), we arrive at the second main conclusion of
this paper, which is an immediate consequence of the first.
\begin{The}
 The Poisson bracket associated with the symplectic form $\Omega$ on covariant
 phase space as defined by Crnkovi\'c, Witten and Zuckerman, according to
 the standard prescription of symplectic geometry, suitably adapted to the
 infinite-dimensional setting encountered in this context, is precisely the
 field theoretical bracket first proposed by Peierls and brought into a more
 geometric form by DeWitt\/:
\begin{equation} \label{eq:POISB4}
 \{\mathslf{F},\mathslf{G}\}[\phi]~
 =~\int_M d^{\,n} x \int_M d^{\,n} y~
   \frac{\delta \mathslf{F}}{\delta\phi^{\,k}}[\phi](x) \;
   G_\phi^{kl}(x,y) \;
   \frac{\delta \mathslf{G}}{\delta\phi^{\,l}}[\phi](y)~.
\end{equation}
\end{The}
Of course, for the expressions in eqs~(\ref{eq:POISB2})-(\ref{eq:POISB4})
to exist, it is not sufficient to require $\mathslf{F}$ and/or $\mathslf{G}\,$
to be local in time. In fact, if we want to use conditions that (a) are
sufficient to guarantee existence of this Poisson bracket without making
use of specific regularity and support properties of the propagator, (b)
are the same for $\mathslf{F}$ and $\mathslf{G}\,$ and (c) are reproduced
under the Poisson bracket, we are forced to impose quite rigid assumptions:
the functionals under consideration must be assumed to be both regular and
local, in the sense that their functional derivative at any point $\phi$
of~$\mathscr{S}$ must be a smooth section of~$V_\phi^{\oast}$ of compact
support (this will force the corresponding Hamiltonian vector field to
be a smooth section of~$V_\phi$ of spatially compact support).

On the other hand, it must be pointed out that this Poisson bracket, which we
might call the Peierls\,-\,DeWitt bracket, has all the structural properties
expected from a good Poisson bracket: bilinearity, antisymmetry, validity of
the Jacobi identity and validity of the Leibniz rule with respect to plain
and ordinary multiplication of functionals. This can be seen directly by
noting that the first two properties and the Leibniz rule are obvious, while
the Jacobi identity expresses the propagator identity for the causal Green
function. But it is of course much simpler to argue that all these properties
follow immediately from the above theorem, in combination with standard results
of symplectic geometry. Moreover, the Peierls\,-\,DeWitt bracket trivially
satisfies the fundamental axiom of field theoretic locality: functionals
localized in spacelike separated regions commute. All this suggests that the
Peierls\,-\,DeWitt bracket is the correct classical limit of the commutator
of quantum field theory. Therefore, it ought to play an outstanding role in
any attempt at quantizing classical field theories through algebraic methods,
a popular example of which is deformation quantization.

The basic complication inherent in the algebraic structure provided by the
Peierls\,-\,DeWitt bracket is that it is inherently dynamical: the bracket
between two functionals depends on the underlying dynamics. This could not
be otherwise. In fact, it is the price to be paid for being able to extend
the canonical commutation relations of classical field theory, representing
a non-dynamical equal-time Poisson bracket, to a covariant Poisson bracket.
The dynamical nature of covariant Poisson brackets is simplified (but still
not trivial) for free field theories, where the equations of motion are
linear, implying that the Jacobi operator $\mathscr{J}[\phi]$ and its
causal Green function $G_\phi$ do not depend on the background solution
$\phi$.

Finally, we would like to remark that the main mathematical condition to
be imposed in order for the constructions presented here to work is that
linearization of the field equations around any solution $\phi$ should
provide a hyperbolic system of partial differential equations on~$M$,
for which existence and uniqueness of the Green functions $G_\phi^\pm$
and $G_\phi$ can be guaranteed. There are various definitions of the concept
of a hyperbolic system that can be found in the literature, but the most
appropriate one seems to be that of regular hyperbolicity, proposed by
Christodoulou \cite{Ch1,Ch2,Ch3} in the context of Lagrangian systems,
according to which the matrix
\[
 u^\mu \, u^\nu \, \frac{\partial^{\>\!2} L}
                        {\partial q_\mu^{i\vphantom{j}} \, \partial q_\nu^j}
\]
should (in our sign convention for the metric tensor) be positive definite
for timelike vectors $u$ and negative definite for spacelike vectors $u$:
a typical example is provided by the sigma model as discussed at the
beginning of this subsection.
What is missing is to translate this condition into the Hamiltonian
formalism and to compare it with other definitions of hyperbolicity
for first order systems, such as the traditional one of Friedrichs.

\section{Conclusions and Outlook}

The approach to the formulation of geometric field theory adopted in
this paper closely follows the spirit of Ref.~\cite{KS1}, in the sense
of emphasizing the importance of combining \linebreak techniques from
multisymplectic geometry with a functional approach. The main novel\-ties
are (a)~the systematic extension from a Lagrangian to a Hamiltonian point
of view, preparing the ground for the treatment of field theories which
have a phase space but no configuration space (or better, a phase bundle
but no configuration bundle), \linebreak (b)~a clearcut distinction between
ordinary and extended multiphase space, which is necessary for a correct
definition of the concept of the covariant Hamiltonian and (c)~the use
of the causal Green function for the linearized operator as the main
tool for finding an explicit formula for the Hamiltonian vector field
associated with a given functional on covariant phase space. This
explicit formula, together with the resulting identification of the
canonical Poisson bracket derived from the standard symplectic form
on covariant phase space with the Peierls\,-\,DeWitt bracket of
classical field theory, are the central results of this paper.

An interesting question that arises naturally concerns the relation
between the Peierls\,-\,DeWitt bracket as constructed here with other
proposals for Poisson \mbox{brackets} in multisymplectic geometry.
In general the latter just apply to certain special classes of functionals.
One such class is obtained by using fields to pull differential forms $f$ 
on extended multiphase space back to space-time and then integrate over 
submanifolds $\Sigma$ of the corresponding dimension. Explicitly, in the 
Lagrangian framework,
\begin{equation} \label{eq:FLAG1}
 \mathslf{F}\,[\phi]~
 =~\int_\Sigma \big( \mathbb{F} \mathscr{L} \smcirc
                     (\varphi,\partial\varphi) \big)^\ast f~,
\end{equation}
whereas in the Hamiltonian framework,
\begin{equation} \label{eq:FHAM1}
 \mathslf{F}\,[\phi]~
 =~\int_\Sigma \big( \mathscr{H} \smcirc (\varphi,\pi) \big)^\ast f~.
\end{equation}
For the particular case of differential forms $f$ of degree $n-1$
and Cauchy hypersurfaces as integration domains $\Sigma$, this kind
of functional was already considered in the 1970's under the name
``local observable''~\cite{Ki} (though on ordinary instead of
extended multi\-phase space), but it was soon noticed that due to
additional restrictions imposed on the forms $f$ allowed in the
construction, the class of functionals so defined is way too small
to be of much use for purposes such as quantization. One of these
restrictions is that $f$ should be what is nowadays called a Hamiltonian 
form~\cite{Ka}. Briefly, an $(n-1)$-form $f$ on~$J^{1\ostar} E$ is said
to be a Hamiltonian form if there exists a (necessarily unique) vector
field $X_f^{}$ on~$J^{1\ostar} E$, called the Hamiltonian vector field
associated with $f$, such that
\begin{equation} \label{eq:HAMVF4}
 \mathrm{i}_{X_f^{}}^{} \omega~=~df~.
\end{equation}
What seems to have motivated this restriction is the possibility to use
the multi\-symplectic analogue of the standard definition~(\ref{eq:POISB1})
of Poisson brackets in mechanics for defining the Poisson bracket between
the corresponding functionals~\cite{KS1}. However, it turns out that, in
contrast to mechanics where $f$ is simply a function, the validity of
eq.~(\ref{eq:HAMVF4}) imposes strong constraints not only on the vector
field $X_f^{}$ but also on the form $f$; in particular, it restricts the
coefficients both of~$X_f^{}$ and of~$f$ in adapted local coordinates to
be affine functions of the multimomentum variables $p\>\!_i^\mu$ and the
energy variable $p\,$~\cite{FR}. (See Refs~\cite{FPR1,FPR2} for a detailed
analysis of the general situation encountered when dealing with the same
question for forms of arbitrary degree.) This implies that the class of
functionals $\mathslf{F}$ derived from Hamiltonian $(n-1)$-forms $f$
according to eqs~(\ref{eq:FLAG1}) and/or~(\ref{eq:FHAM1}) does not
close under ordinary multiplication of functionals.
 
Fortunately, using the Peierls\,-\,DeWitt bracket between functionals,
we may dispense with the restriction to Hamiltonian forms. In fact,
this line of reasoning was already followed by the authors of Ref.~%
\cite{KS1},  where both the symplectic form on the solution space and
the corresponding Poisson bracket between functionals on the solution
space, with all its structurally desirable properties, are introduced
explicitly. What remained unnoticed at the time was that this bracket
is just the Peierls\,-\,DeWitt bracket of physics and that incorporating
the theory of ``local observables'' into this general framework results
in the transformation of a definition, as given in Ref.~\cite{Ki}, into
a theorem which, in modern language, states that the Peierls\,-\,DeWitt
bracket $\{\mathslf{F},\mathslf{G}\,\}$ between two functionals
$\mathslf{F}$ and $\mathslf{G}\,$ derived from Hamiltonian $(n-1)$-forms
$f$ and~$g$, respectively, is the functional derived from the Hamiltonian
$(n-1)$-form $\{f,g\}$. An explicit proof, based on the classification
of Hamiltonian vector fields and Hamiltonian $(n-1)$-forms that follows
from the results of Ref.~\cite{FPR2}, has been given recently~\cite{Sa};
details will be published elsewhere.

Of course, there is a priori no reason for restricting this kind of
investigation to forms of degree $n-1$, since physics is full of
functionals that are localized on submanifolds of space-time of
other dimensions, such as: values of observable fields at space-time
points (dimension $0$), Wilson loops (traces of parallel transport
operators around loops) in gauge theories (dimension $1$), etc..
This problem is presently under investigation.

\section*{Appendix: Affine Spaces and Duality}

In this appendix, we collect some basic facts of linear algebra for
affine spaces which are needed in this paper but which do not seem
to be readily available in the literature.
 
A (nonempty) set $A$ is said to be an \emph{affine space} modelled
on a vector space $V$ if there is given a map
\begin{equation} \label{eq:sum}
 \begin{array}{ccccc}
  + &:& A \times V & \longrightarrow &  A \\
    & &    (a,v)   &   \longmapsto   & a+v 
 \end{array}
\end{equation}
satisfying the following two conditions:
\begin{itemize}
 \item $a+(u+v)=(a+u)+v \,$ for all $a \in A$ and all $u,v \in V$.
 \item Given $a,b \in A$, there exists a unique $v \in V$ such that
       $a=b+v$.
\end{itemize}
Elements of $A$ are called points and elements of $V$ are called vectors,
so the map (\ref{eq:sum}) can be viewed as a transitive and fixed point
free action of $V$ (as an Abelian group) on $A$, associating to any point
and any vector a new point called their sum. Correspondingly, the vector
$v$ whose uniqueness and existence is postulated in the second condition
is often denoted by $a-b$ and called the difference of the points $a$
and $b$.

For every affine space $A$, the vector space on which it is modelled
is determined uniquely up to isomorphism and will usually be denoted
by $\vec{A}$.
 
A map $\, f: A \rightarrow B \,$ between affine spaces $A$ and $B$ is
said to be \emph{affine} if there exists a point $a \in A$ such that
the map $\, \vec{f}_a: \vec{A} \rightarrow \vec{B} \,$ defined by
\begin{equation}
 \vec{f}_a(v)~=~f(a+v)-f(a)
\end{equation}
is linear, that is, $\vec{f}_a \in L(\vec{A},\vec{B})$. It is easily
seen that this condition does not depend on the choice of the reference
point: in fact, if the map $\vec{f}_a$ is linear for some choice of~$a$,
then the maps $\vec{f}_{a'}$ are all equal as $a'$ varies through $A$,
so it makes sense to speak of the \emph{linear part} $\vec{f}$ of an
affine map $f$. Denoting the set of all affine maps from $A$ to $B$
by $A(A,B)$, we thus have a projection
\begin{equation} \label{eq:lpaff}
 \begin{array}{ccccc}
  l &:& A(A,B) & \longrightarrow & L(\vec{A},\vec{B}) \\
    & &   f    &   \longmapsto   &      \vec{f} 
 \end{array}~.
\end{equation}
This construction is particularly important in the special case where
$B$ is itself a vector space, rather than just an affine space. Given
an affine space $A$ and a vector space $W$, the set $A(A,W)$ of affine
maps from $A$ to $W$ is easily seen to be a vector space: in fact it
is simply a linear subspace of the vector space $\Map(A,W)$ of all
maps from $A$ to $W$. Moreover, the projection
\begin{equation} \label{eq:lpvec}
 \begin{array}{ccccc}
  l &:&A(A,W) & \longrightarrow & L(\vec{A},W) \\
    & &  f    &   \longmapsto   &    \vec{f} 
 \end{array}
\end{equation}
is a linear map whose kernel consists of the constant maps from $A$
to $W$. Identifying these with the elements of $W$ itself, we obtain
a natural isomorphism
\begin{equation}
 A(A,W)/W~\cong~L(\vec{A},W)~,
\end{equation}
or equivalently, an exact sequence of vector spaces, as follows:
\begin{equation} \label{eq:esaff1}
 0~\longrightarrow~W~\longrightarrow~A(A,W)~
   \stackrel{l}{\longrightarrow}~L(\vec{A},W)~\longrightarrow~0~.
\end{equation}
In the general case, one shows that given two affine spaces $A$ and $B$,
the set $A(A,B)$ of affine maps from $A$ to $B$ is again an affine space,
such that $\overrightarrow{A(A,B)\;} = A(A,\vec{B})$, and that the
projection (\ref{eq:lpaff}) is an affine map.

Concerning dimensions, we may choose a reference point $o$ in $A$ which
provides not only an isomorphism between $A$ and $\vec{A}$ but also a
splitting of the exact sequence (\ref{eq:esaff1}) and hence an isomorphism
between $A(A,W)$ and $\, W \oplus L(\vec{A},W)$, to show that
\begin{equation} \label{eq:dimaff1}
 \dim A(A,W)~=~\dim W + \, \dim L(\vec{A},W)~.
\end{equation}

Choosing $W$ to be the real line $\R$, we obtain the \emph{affine dual}
$A^\star$ of an affine space~$A$:
\begin{equation} \label{eq:oduaff}
 A^\star~=~A(A,\R)~.
\end{equation}
Observe that this is not only an affine space but even a vector space
which, according to eq.~(\ref{eq:esaff1}), is a one-dimensional extension
of the linear dual $\vec{A}^\ast$ of the model space $\vec{A}$ by~$\R$,
that is, we have the following exact sequence of vector spaces:
\begin{equation} \label{eq:esaff2}
 0~\longrightarrow~\R~\longrightarrow~A^\star~
   \stackrel{l}{\longrightarrow}~\vec{A}^\ast~\longrightarrow~0~.
\end{equation}
In particular, according to eq.~(\ref{eq:dimaff1}), its dimension equals 1
plus the dimension of the original affine space:
\begin{equation} \label{eq:dimaff2}
 \dim A^\star~=~\dim A + 1~.
\end{equation}
More generally, we may replace the real line $\R$ by a (fixed but arbitrary)
one-dimensional real vector space $R$ (which is of course isomorphic but in
general not canonically isomorphic to $\R$) to define the \emph{twisted
affine dual} $A^{\ostar}$ of an affine space $A$:
\begin{equation} \label{eq:tduaff}
 A^{\ostar}~=~A(A,R)~.
\end{equation}
Again, this is not only an affine space but even a vector space which,
according to eq.~(\ref{eq:esaff1}), is a one-dimensional extension of
the linear dual $\vec{A}^\ast$ of the model space $\vec{A}$ by~$R$,
that is, we have the following exact sequence of vector spaces:
\begin{equation} \label{eq:esaff3}
 0~\longrightarrow~R~\longrightarrow~A^{\ostar}~
   \stackrel{l}{\longrightarrow}~\vec{A}^{\oast}~\longrightarrow~0~.
\end{equation}
Obviously, the dimension is unchanged:
\begin{equation} \label{eq:dimaff3}
 \dim A^{\ostar}~=~\dim A + 1~.
\end{equation}
Moreover, we have the following canonical isomorphism of vector spaces
\begin{equation} \label{eq:otdaff}
 A^{\ostar}~\cong~A^\star \otimes R~,
\end{equation}
and more generally, for any vector space $W$,
\begin{equation} \label{eq:daff}
 A(A,W)~\cong~A^\star \otimes W~.
\end{equation}
Regarding the splittings of the exact sequence (\ref{eq:esaff3}), we note
the following fact which is used in the construction of the inverse Legendre
transformation: these splittings form an affine space modelled on the bidual
$\vec{A}^{\ast\ast}$ of $\vec{A}$, which in finite dimensions can be
identified with $\vec{A}$ itself.

The concept of duality applies not only to spaces but also to maps between
spaces: given an affine map $\, f : A \rightarrow B \,$ between affine spaces
$A$ and $B$, the formula
\begin{equation} \label{eq:duaffm}
 (f^\star(b^\star))(a)~=~b^\star(f(a)) \qquad
 \mbox{for $\, b^\star \smin B^\star, a \smin A$}
\end{equation}
yields a linear map $\, f^\star : B^\star \rightarrow A^\star \,$ between
their affine duals $B^\star$ and $A^\star$. As a result, the operation of
taking the affine dual can be regarded as a (contravariant) functor from
the category of affine spaces to the category of vector spaces. This functor
is compatible with the usual (contravariant) functor of taking linear duals
within the category of vector spaces in the sense that the following diagram
commutes:
\begin{equation}
 \begin{array}{ccc}
     B^\star   & \stackrel{f^\star}{\longrightarrow} &
     A^\star     \\[2mm]
  \downarrow &                                       &
  \downarrow     \\[1mm]
  \vec{B}^\ast & \stackrel{\vec{f}^\ast}{\longrightarrow} &
  \vec{A}^\ast
 \end{array}
\end{equation}

Finally, we also need the construction of quotients in the affine
category. These are defined by dividing out not affine subspaces but
rather linear subspaces of the model space. In fact, given an affine
space $A$ and a linear subspace $V$ of its model space $\vec{A}$, we
can declare two points $a$ and $a'$ of $A$ to be \emph{equivalent}
modulo $V$ if $\, a - a' \in V$. Obviously, this relation is reflexive,
symmetric and transitive, and hence is an equivalence relation dividing
$A$ into equivalence classes; the set of equivalence classes is as usual
denoted by $A/V$. It is then easy to see that there is a unique affine
structure on $A/V$ turning $A/V$ into an affine space such that
$\, \overrightarrow{A/V\,\>} = \vec{A}/V$ and such that the natural
projection
\begin{equation}
 \begin{array}{ccccc} 
  \rho &:& A & \longrightarrow & A/V \\
       & & a &   \longmapsto   & [a]
 \end{array}
\end{equation}
is an affine map. Moreover, this construction satisfies the standard
factorization property: given two affine spaces $A$ and $B$, two linear
subspaces $V$ of $\vec{A}$ and $W$ of $\vec{B}$ and an affine map
$\, f : A \longrightarrow B \,$ whose linear part $\, \vec{f} :
\vec{A} \longrightarrow \vec{B} \,$ maps $V$ into $W$, there
exists a unique affine map $\, [f] : A/V \longrightarrow B/W \,$
such that the diagram
\begin{equation}
 \begin{array}{rcccl}
                &          A      &
  \stackrel{\rule[-2mm]{0mm}{2mm} {\textstyle f}
            \rule[-2mm]{0mm}{2mm}}{\longrightarrow} & B & \\[3mm]
  \rho \!\!\!\!& \bigg\downarrow & & \bigg\downarrow &\!\!\!\! \rho \\
                &         A/V     &
  \stackrel{\rule[-2mm]{0mm}{2mm} {\textstyle [f]}
            \rule[-2mm]{0mm}{2mm}}{\longrightarrow} & B/W &
 \end{array}
\end{equation}
commutes.

Concluding this appendix, we would like to point out that all the concepts
introduced above can be extended naturally from the purely algebraic setting
to that of fiber bundles. For example, affine bundles are fiber bundles
modelled on an affine space whose transition functions (with respect to
a suitably chosen atlas) are affine maps. Moreover, functors such as the
affine dual or the construction of quotients are smooth (see \cite{La}
for a definition of the concept of smooth functors in a similar context)
and therefore extend naturally to bundles (over a fixed base manifold $M$).
This means that any affine bundle $A$ over $M$ has a naturally affine dual,
which is a vector bundle $A^\star$ over $M$, and that given any vector
subbundle $V$ of the difference vector bundle $\vec{A}$ of an affine
bundle $A$ over $M$, we can form the quotient affine bundle $A/V$
over $M$.

\pagebreak

\end{document}